\documentclass[twocolumn,pre,floatfix]{revtex4}
\usepackage{psfrag,epsfig,amsfonts,amssymb,amsmath,wasysym,bm}
\usepackage{dcolumn}
\usepackage{bbold}
\usepackage[normalem]{ulem}
\usepackage{color}
\usepackage{hyperref}

\newcommand{\Imc}{I_{\mathrm{mc}}}

\newcommand{\pr}{{\rm{Prob}}}

\newcommand{\bw}{h_{\rm{BW}}}

\newcommand{\RR}{{\mathbb R}}

\newcommand{\NN}{{\mathbb N}}

\newcommand{\Aeq}{{\cal A}_{\!\!\;\rm{eq}}}

\newcommand{\trel}{t_{\mathrm{rel}}}

\newcommand{\propa}{{\cal U}}
\newcommand{\Aobs}{{\cal A}}

\newcommand{\tr}{\mbox{Tr}}

\newcommand{\rhomic}{\rho^0_{\mathrm{mc}}}

\providecommand{\norm}[1]{\|#1\|}
\newcommand{\da}{\Delta_{\! A}}

\newcommand{\amax}{a_{\mathrm{max}}}
\newcommand{\amin}{a_{\mathrm{min}}}
\newcommand{\hh}{\hbar}

\begin{document}

\title{Typicality of Prethermalization}
\author{Peter Reimann and Lennart Dabelow}
\affiliation{Fakult\"at f\"ur Physik, 
Universit\"at Bielefeld, 
33615 Bielefeld, Germany}

\date{\today}

\begin{abstract}
Prethermalization refers to the remarkable
relaxation behavior which an integrable 
many-body system in the presence of a weak 
integrability-breaking perturbation may exhibit:
After initial transients have died out,
it stays for a long time close to some 
non-thermal steady state,
but on even much larger time scales 
it ultimately switches over to the
proper thermal equilibrium behavior.
By extending Deutsch's conceptual framework
from Phys.~Rev.~A~{\bf 43}, 2046 (1991),
we analytically predict that prethermalization
is a typical feature for a very general class of
such weakly perturbed systems.
\end{abstract}

\maketitle

Isolated many-body quantum
systems are known to 
equilibrate,
i.e., expectation values 
exhibit
an initial relaxation and then
spend most of their time close 
to a constant value,
provided 
some rather weak preconditions
are fulfilled.
Furthermore, thermalization is expected
for so-called non-integrable 
systems,
i.e., the long-time behavior is well approximated by 
a
microcanonical ensemble.
(Possible exceptions, e.g., due to 
many-body localization, are tacitly 
ignored here.)
In contrast, integrable systems usually
exhibit
quite significant deviations from such 
a thermal long-time behavior.
All these issues have been extensively 
explored in the literature,
as reviewed, among others, in Refs.~\cite{gog16,ale16,bor16,mor18}.
They are not the subject of our present 
work but rather will be taken for granted.

Our main issue 
is the question of how the temporal 
relaxation of an integrable system changes in 
response to a weak integrability-breaking 
perturbation.
More specifically,
we will derive a rigorous bound
for the difference between unperturbed 
and perturbed expectation values,
implying that those changes 
remain over a long period of time
negligibly small 
for a very large class of 
weak perturbations.
Our approach is conceptually akin 
to Deutsch's seminal work 
on thermalization,
treating the perturbations 
along the lines of
random matrix theory \cite{deu91}.
In particular, we will exploit
Deutsch's result concerning
the ultimate thermalization
of the perturbed systems.
With respect to the unperturbed (integrable) 
system, we will moreover 
take for granted that its initial 
relaxation is not extremely 
slow, and 
that it exhibits clearly 
observable deviations from a 
thermal long-time behavior.
Altogether, we are thus left with a 
very large class of 
perturbations with the following quite remarkable
property, henceforth 
named prethermalization:
Initially, the perturbed system closely follows 
the unperturbed relaxation towards a non-thermal 
steady state, but on even much larger time 
scales, there must be a clearly visible
transition to the ultimate thermal 
behavior.

Originally, the term prethermalization 
was introduced by Berges, Bors\'anyi, and 
Wetterich \cite{ber04} 
for 
matter under extreme conditions in a
quasi-steady state far from equilibrium, 
which nevertheless exhibits some genuine 
thermal properties, 
however without any reference to the 
concept of integrability.
Our 
present,
somewhat different
notion of prethermalization
has been independently established 
by Moeckel and Kehrein in 
Ref.~\cite{moe08}.
During recent years, these and further 
slightly differing guises of 
prethermalization have been 
explored in numerous theoretical 
\cite{div1,div2a,div2b,ber16}
as well as experimental 
\cite{div3} investigations,
see also the recent reviews 
\cite{lan16,mor18} and further
references therein.

Incidentally, the particular examples
in Moeckel and Kehrein's original work 
\cite{moe08} and also in some
subsequent studies \cite{div2a}
are beyond the above mentioned
realm of our present approach: 
If the unperturbed system is initially 
at thermal equilibrium or in
the energy ground state, 
as it is the case in \cite{moe08,div2a},
then the unperturbed dynamics
is trivial, and also the signatures of 
prethermalization after adding a weak 
perturbation remain too small for our 
purposes.

Against our treating the perturbations as random 
matrices (in the unperturbed energy basis), 
one might object that the ``true'' perturbation 
in any concrete physical model is not a random matrix.
In particular, 
the true matrix is often 
banded 
\cite{deu91,fei89,fyo96,gen12},
i.e., the typical magnitude of 
its entries
decreases with increasing distance from the 
matrix diagonal.
Furthermore, 
for non-interacting systems perturbed by 
few-body interactions,
the matrix will be very sparse, i.e.,
only a small fraction of 
its entries
is non-zero \cite{bor16,bro81,fyo96,fla97}.

To overcome these concerns,
we will 
consider 
ensembles of random matrices which can 
be tailored
to emulate the basic features of many 
concrete models,
such as sparsity, bandedness, and other 
statistical characteristics
\cite{bor16,deu91,fei89,fyo96,gen12,bro81,fla97}.
The true perturbation 
is thus expected to be contained as one 
specific matrix in 
such a properly tailored ensemble as well.
(For simplicity, one may imagine matrices
of large but finite dimension, whose 
entries
can assume only a finite number of different 
possible values, as it is the case in any 
numerical investigation. If each possible value
has non-zero probability, there is a finite chance 
to sample the true matrix from the ensemble.)
Hence, if one could prove that some property 
applies to all 
members
of the ensemble, the
property would also apply to the true model.
Our main result consists in a slightly weaker statement, 
namely that the property ``prethermalization'' 
at least applies 
with overwhelming probability
when randomly sampling perturbations from the 
ensemble (``typicality of prethermalization'').
It is therefore still very reasonable to expect 
that the true model is not one of the
extremely unlikely exceptions.
An illustrative example (spin chain 
model) is provided in the 
Supplemental Material \cite{sup}.
Analogous arguments are routinely adopted
in random matrix theory, which is well
known to be extremely successful in 
practice \cite{bor16,bro81},
though its applicability has to our
knowledge not been rigorously justified in 
any concrete physical example.
Similar considerations also
apply to many other 
``non-systematic''
but practically very well established 
approximations, such as density functional 
theory or Boltzmann equations beyond
the validity limits of their derivation.

We will demonstrate typicality of prethermalization
for a great variety of different ensembles.
The resulting total set of all admitted 
perturbations is therefore extremely large.
This seems to us a quite noteworthy finding
in itself, independent of the question
whether some particular model is 
covered or not.
Moreover, 
to actually exclude 
some particular model, it 
would have
to be untypical
with respect to every one
of those various ensembles.
We finally remark that most applications 
of random matrix theory focus on the
ensemble-averaged behavior and
take for granted that most individual matrices 
behave very similarly to the average
\cite{bor16,bro81}.
In our present approach, no such
extra assumption will be needed.

The unperturbed system is described by a Hamiltonian
$H_0$ with eigenvalues $E^0_n$ and eigenvectors
$|n\rangle_{\! 0}$.
The unperturbed evolution of an arbitrary 
initial state
$\rho(0)$ can thus be written as
$\rho_0(t)=e^{-iH_0t/\hbar}\rho(0)e^{iH_0t/\hbar}$
and the
expectation value of any given observable 
$A$ 
as
\begin{eqnarray}
& & \!\!\!\! \!\!\!\!
\Aobs_0(t) 
:= 
\tr\{\rho_0(t)A\} = \sum_{mn} \rho^0_{mn}(0)\, A^0_{nm}\, 
e^{i\frac{E^0_n-E^0_m}{\hbar}t}
\, , \ \ \
\label{10}
\\
& & \!\!\!\!  \!\!\!\!
\rho^0_{mn}(t) 
:=
\, _0\!\langle m | \rho_0(t) | n\rangle_{\! 0}\ ,\ \ 
A^0_{nm} :=\, _0\!\langle n | A | m \rangle_{\! 0}
\ ,
\label{20}
\end{eqnarray}
where, depending on the specific system under consideration, 
the indices $m$ and $n$ run from $1$ 
to infinity or to some finite upper limit.

Likewise, the perturbed system
\begin{eqnarray}
H = H_0 +V
\label{30}
\end{eqnarray}
exhibits
eigenvalues $E_n$ and eigenvectors $|n\rangle$.
Focusing on the same initial state $\rho(0)$ as before,
the expectation value $\Aobs (t)$ under the perturbed dynamics
is then given by the same formulas as in (\ref{10}) 
and (\ref{20}), except that all indices ``$0$'' 
must be omitted.
In terms of the 
unitary basis transformation matrix
\begin{eqnarray}
U_{mn}:=\langle m|n\rangle_{\!0}
\ ,
\label{40}
\end{eqnarray}
this expectation value 
can be further rewritten as
\begin{eqnarray}
\Aobs(t) =
\!
\sum_{mn}
\!
\sum_{\mu\nu\sigma\tau} 
\!
U_{m\mu}U^\ast_{n\nu}U_{n\sigma}U^\ast_{m\tau}
\rho^0_{\mu\nu}(0)
A^0_{\sigma\tau} 
e^{i\frac{E_n-E_m}{\hbar}t}
\, .\,
\label{50}
\end{eqnarray}

The quantity of foremost interest is the difference 
\begin{eqnarray}
\Delta(t):= \Aobs(t)-\Aobs_0(t)
\label{60}
\end{eqnarray}
between
the perturbed and the unperturbed expectation values.
Taking into account $\rho^0_{mn}(0)e^{i(E^0_n-E^0_m)t/\hbar}=\rho^0_{mn}(t)$
(see above (\ref{10})),
it follows with (\ref{10}) and (\ref{50}) that
\begin{eqnarray}
\Delta(t) & = &  \sum_{\mu\nu\sigma\tau} \rho^0_{\mu\nu}(t)A^0_{\sigma\tau}\, 
[\gamma_{\tau\mu}(t) \gamma^\ast_{\sigma\nu}(t)- \delta_{\tau\mu}\delta_{\sigma\nu}]
\, ,
\label{70}
\\
\gamma_{\tau\mu}(t) 
& := & 
\sum_{m} 
U^\ast_{m\tau}U_{m\mu}\,
e^{i(E_\mu^0-E_m)t/\hbar}
\ ,
\label{80}
\end{eqnarray}
where $\delta_{mn}$ is the Kronecker delta.

Finally, instead of one particular perturbation 
$V$ in (\ref{30}), we consider a statistical 
ensemble of different $V$'s, and we indicate
averages
over the ensemble
by an overline.
This randomization of $V$ is inherited
by the Hamiltonian  $H$ in (\ref{30}) 
and thus by the eigenvalues
$E_n$, the eigenvectors $|n\rangle$,
the $U_{mn}$ in (\ref{40}), 
and the $\gamma_{\tau\mu}(t)$ 
in (\ref{80}).
On the other hand, $H_0$, $\rho(0)$, 
and $A$ are considered as arbitrary
but fixed (non-random), hence the same 
must apply to $E_n^0$, $|n\rangle_{\! 0}$, 
$\rho_0(t)$, and to the matrix elements 
in (\ref{20}).

The first main result of our Letter consists in 
the general rigorous bound
\begin{eqnarray}
\overline{|\Delta(t)|} 
& \leq & 
\frac{\da}{2}\, f(t)
\ ,
\label{90}
\\
f(t)
& := &
3\sqrt{1-Y(t)}+\sqrt{1-Y(t)+W(t)}
\ ,
\label{100}
\\
Y(t)
& := & 
\sum_{\mu\nu\sigma} 
\rho^0_{\mu\nu}(t)\,
\overline{\gamma_{\sigma\mu}(t)}\,
\left[\overline{\gamma_{\sigma\nu}(t)}\right]^\ast
\ , 
\label{110}
\\
W(t) 
& := & 
4Y(t) -[Z(t)+Z^\ast(t)]^2
\ ,
\label{120}
\\[0.2cm]
Z(t)
& := & 
\sum_{\mu\nu} 
\rho^0_{\mu\nu}(t)\,
\overline{\gamma_{\nu\mu}(t)}
\ ,
\label{130}
\end{eqnarray}
where $\da$ is the measurement range of $A$
(largest minus smallest eigenvalue).
The quite tedious derivation
has been relegated to the 
Supplemental Material \cite{sup}.

Applying Markov's inequality 
to (\ref{90}),
it follows for any $\epsilon>0$ 
that
\begin{eqnarray}
\pr\big(\,
|\Delta (t)| \leq \epsilon\,\da \big)
\geq 1-  f(t)/2\epsilon
\ ,
\label{140}
\end{eqnarray}
where the left hand side denotes the probability
that $|\Delta (t)| \leq \epsilon\,\da$
when randomly sampling perturbations $V$.
For sufficiently small $f(t)$, the 
difference $\Delta(t)$ in (\ref{60}) will thus be 
negligible for the vast majority of all $V$'s.

Our first assumption regarding the 
so far arbitrary 
ensemble of $V$'s is as follows:
Multiplying the unperturbed energy
eigenvectors $|n\rangle_{\!0}$ by
arbitrary factors 
$\sigma_n\in\{\pm1\}$ 
leaves 
the $V$ ensemble invariant.
Hence also the statistical properties of 
$\gamma_{\mu\nu}$ in (\ref{180}) 
remain unchanged if all the matrix 
elements $U_{mn}$ in (\ref{40}) 
are multiplied by arbitrary 
factors $\sigma_n \in\{\pm 1\}$.
As a consequence (see also \cite{sup}), 
the ensemble average 
of (\ref{80}) must vanish unless 
$\tau=\mu$,
\begin{eqnarray}
\overline{\gamma_{\tau\mu}(t)}
& = & 
\delta_{\tau\mu}\, \overline{g_\mu(t)}
\ ,
\label{150}
\\
g_\mu(t) & := & \sum_{m} |U_{m\mu}|^2\,
e^{i(E_\mu^0-E_m)t/\hbar}
\label{160}
\ .
\end{eqnarray}
To justify this assumption we 
note that randomly flipping the signs of the 
$|n\rangle_{\!0}$ 
leaves all physical properties unchanged 
but randomizes the signs
of the true perturbation matrix 
elements $V_{mn}^0$.
Hence, it is appropriate to adopt a
random matrix model
with the above invariance property.

As stated in the introduction,
the unperturbed system is assumed to 
exhibit equilibration but not thermalization. 
Implicitly, this requires a 
macroscopically well defined 
system energy; i.e.,
there must exist a microcanonical energy 
interval $\Imc :=[E-\delta E,E]$ so that 
only energies $E^0_n\in \Imc $ exhibit 
non-negligible level populations
$\rho^0_{nn}(0)$.
The number of energies $E_n^0$ contained
in $\Imc $ is denoted by $N$ and, without loss of 
generality, we assume that $n\in\{1,...,N\}$ 
for all those $E_n^0$'s.
Furthermore, whenever  $E_n^0\not\in \Imc $,
we adopt the idealization that 
$\rho^0_{nn}(0)$ is 
strictly zero \cite{f1}.
The Cauchy-Schwarz inequality
$|\rho^0_{mn}(0)|^2\leq \rho^0_{mm}(0) \rho^0_{nn}(0)$
then implies that in (\ref{10})
only summands with $m,n\in\{1,...,N\}$ 
actually contribute.
As usual, 
we take for granted
that $N$ is huge (exponentially large 
in the system's degrees of 
freedom \cite{gol10a}), while the 
local level density 
remains close to 
$D:=\delta E/N$ throughout 
the interval $\Imc $.

Given that only indices $m,n\in\{1,...,N\}$ actually 
matter in (\ref{10}), we can and will assume that
their range is
extended to arbitrary integer values
and  that the energies $E_n^0$ 
and the matrix elements 
$V^0_{mn} :=\, _0\!\langle m | V | n \rangle_{\! 0}$ 
are (re-)defined for arbitrary 
integers $m,n\not\in\{1,...,N\}$ by way 
of ``extrapolating'' in a physically natural 
way their properties for
$m,n\in\{1,...,N\}$.

As a first example, we consider the 
particularly simple case that
$E^0_{n+1}-E^0_n=D$ for all $n$, and that
the statistical properties 
of the matrix elements 
$V^0_{mn}$
do not depend separately on $m$ and $n$, 
but only on the difference $m-n$.
As a consequence (see also \cite{sup}),
the statistical properties of (\ref{160}) remain invariant 
when simultaneously adding an arbitrary integer 
$\nu$ to all indices on the right hand side
(but not on the left hand side).
Upon averaging, one can thus infer that
$\overline{g_\mu(t)}=\overline{g_{\mu+\nu}(t)}$,
hence
\begin{eqnarray}
g(t):=\overline{g_\mu(t)}
\label{170}
\end{eqnarray}
is a well-defined ($\mu$-independent) 
function.

Under the additional assumption that
all statistical properties of the 
diagonal matrix elements 
$V^0_{nn}$
are identical to those of 
$-V^0_{nn}$, 
one can finally show \cite{sup}
that 
\begin{eqnarray}
g(t)
=[g(t)]^\ast
\ .
\label{180}
\end{eqnarray}
To justify this assumption we note that
the diagonal elements of the true
$V$ in (\ref{30}) can always be 
re-adjusted to vanish on the average.
A symmetrization procedure for the remaining
distribution will be provided later.

Introducing (\ref{150})-(\ref{180}) 
into (\ref{110})-(\ref{130}), and 
taking into account that 
$\sum_\nu\rho^0_{\nu\nu}=1$
yields $Y(t)=[g(t)]^2$, $Z(t)=g(t)$, 
and $W(t)=0$,
hence (\ref{100}) takes the form
\begin{eqnarray}
f(t)=4\sqrt{1-[g(t)]^2}
\ .
\label{190}
\end{eqnarray}

One readily infers from (\ref{40}), (\ref{160}), 
and (\ref{170}) that $g(0)=1$ and that
$|g(t)|\leq 1$ for all $t$.
Furthermore,
it is convenient to rewrite (\ref{160}) as
\begin{eqnarray}
g_\mu(t) 
& = & 
\int dE\, h_\mu(E)\, e^{-i E t/\hbar}
\ ,
\label{200}
\\
h_\mu(E)
& := &
\sum_{m} |U_{m\mu}|^2\,\delta(E-E_m+E_\mu^0)
\ .
\label{210}
\end{eqnarray}
The quantity $F_\mu(E):=h_\mu(E-E_\mu^0)$
plays a key role in 
random matrix theory under the name 
strength function 
or local spectral density of states 
\cite{fyo96,bor16}.
Specifically, one finds that 
the ensemble average
$\overline{h_\mu(E)}$ 
is very well approximated by the
Breit-Wigner distribution
\begin{eqnarray}
\bw (E) :=
\frac{1}{2\pi} \frac{\Gamma}
{E^2+\Gamma^2/4}
\label{220}
\end{eqnarray}
under conditions which, together with 
the concomitant definition of 
$\Gamma$, will be discussed in more 
detail shortly.
Introducing this result into (\ref{200}) 
yields $\overline{g_\mu(t)} =e^{-\Gamma\,|t|/2\hbar}$,
and with (\ref{170}), (\ref{190}) 
we obtain
\begin{eqnarray}
f (t)
= 4 \sqrt{1-e^{-\Gamma |t|/\hbar}}
\leq  4\sqrt{\Gamma |t|/\hbar}
\ .
\label{230}
\end{eqnarray}

Eqs.~(\ref{140}) and (\ref{230}) represent our
main results.
In the remainder of the Letter we 
focus
-- as usual in random matrix theory 
\cite{fyo96,bor16,bro81} --
on the case that all 
$V_{mn}^0$ with $m\geq n$  
are statistically independent of each other 
(those with $m<n$ follow from 
$V^0_{nm}=[V^{0}_{mn}]^\ast$),
that the statistics only depends
on $m-n$ (see above (\ref{170})), and
that $V_{mn}^0$ and $-V_{mn}^0$
are equally likely
(see above Eqs.~(\ref{150}), (\ref{180}) and \cite{sup}).

If all $V_{mn}^0$ 
are furthermore real and 
Gaussian
distributed 
with variance 
$\sigma_{\! v}^2$,
the result (\ref{220}) with
\begin{eqnarray}
\Gamma := 2\pi \sigma_{\! v}^2/D
\label{240}
\end{eqnarray}
was obtained
by Deutsch \cite{deu91}.e
been worked out 
by Fyodorov et al. in Refs.~\cite{fyo96},
including
distributions 
with
a pronounced 
delta peak at zero, 
corresponding to {\em sparse} 
random matrices $V_{mn}^0$.
In addition, they also admitted the possibility
of {\em banded} matrices \cite{f5}.
We have further 
extended their analytical 
supersymmetry approach,
and moreover performed extensive numerical 
explorations,
showing that the key results
(\ref{220})-(\ref{240}) 
remain valid also
for complex 
$V_{mn}^0$'s and under still 
considerably weaker assumptions
regarding 
their statistics.
A few illustrative examples are provided 
in the Supplemental Material \cite{sup}.

Multiplying $V$ in (\ref{30}) 
by an extra factor $\lambda$
(coupling strength) entails a 
factor $\lambda^2$ in (\ref{240}),
hence the characteristic time scale
in (\ref{230}) decreases as 
$\lambda^{-2}$, in
agreement with previous 
findings 
for the persistence of the 
prethermalized state
\cite{mor18,ber16} .
However, we note that our inequality
(\ref{90}) 
admits strictly speaking no conclusions 
regarding the actual appearance of
non-small differences in (\ref{60}).

In order to abandon the requirement
of equally spaced energies $E_n^0$
(see above Eq.~(\ref{170})),
let us consider
an unperturbed 
Hamiltonian $\tilde H_0$ with the same
eigenvectors $|n\rangle_{\! 0}$ as the
original $H_0$, 
but with modified energies
$\tilde E_n^0=E_n^0+\epsilon_n$.
In view of (\ref{10}) one anticipates
that the corresponding expectation value
$\tilde \Aobs_0(t)$ still remains 
close to $\Aobs_0(t)$ for sufficiently small 
$\epsilon_n$ and not too 
large $t$.
Indeed, it can be rigorously shown \cite{sup} that
\begin{eqnarray}
|\tilde \Aobs_0(t)-\Aobs_0(t)|
\leq \da 
\,|t|
\max_{1\leq n \leq N}
|\epsilon_n|
/\hbar
\ .
\label{250}
\end{eqnarray}
Taking for granted that the unperturbed 
Hamiltonian $\tilde H_0$
exhibits equilibration but not thermalization
(see beginning of the Letter), we denote 
by $\trel$ its relaxation time;
i.e., $\tilde \Aobs_0(t)$ 
remains very close to some (non-thermal) 
equilibrium value
$\Aeq$ for (almost) all $t\geq \trel$.
It follows with (\ref{250}) that
also $\Aobs_0(t)$ exhibits 
practically the same initial 
relaxation behavior and then remains
close to $\Aeq$ for quite some time, 
provided $|\epsilon_n|\ll \hbar/\trel$
for all $n=1,...,N$.
Recalling that the energy level
density is exponentially large in the 
degrees of freedom \cite{gol10a},
these conclusions must actually apply 
to rather general non-equidistant
energies $\tilde E_n^0$.

Returning to our perturbed systems of 
the form (\ref{30}), where the considered 
ensemble of $V$'s satisfies the rather
weak assumptions mentioned above,
we can thus conclude from (\ref{60}), 
(\ref{140}), (\ref{230}), and (\ref{240}) 
that also the perturbed expectation 
values $\Aobs(t)$ exhibit an 
initial relaxation and then remain 
close to $\Aeq$ for quite 
some time \cite{f3},
at least for the vast majority of 
perturbations $V$,
and provided they are sufficiently 
weak so that
\begin{eqnarray}
\sigma_{\! v}^2\ll 
 \frac{1}{32\pi}
 \frac{\hbar D}{\trel}
\ .
\label{260}
\end{eqnarray}
On the other hand, ultimate thermalization
for most such $H$'s in (\ref{30})
has been established in
Refs.~\cite{deu91,rei15b}.
Recalling the considerations at the beginning
of our paper, all those ``typical'' 
$H$'s thus exhibit prethermalization.

Finally, upon defining modified perturbations 
$\tilde V$ via $\tilde V^0_{mn} :=V^0_{mn}-\delta_{mn}\epsilon_n$,
we can conclude with Eq.~(\ref{30}) that $\tilde H_0+\tilde V=H$.
Hence, the vast majority of
those perturbations $\tilde V$ of
$\tilde H_0$ must, again, entail prethermalization.
In doing so, the $ \epsilon_n$'s 
are often expected to be so small that 
the resulting ensemble of $\tilde V$'s 
is almost identical to the original ensemble of $V$'s
(see below).
More generally, 
since the modified 
energies $\tilde E_n^0$ 
need no longer be ordered by magnitude,
even substantially more general ensembles 
of $\tilde V$'s than of $V$'s are actually
admitted, see also \cite{sup}.
Along similar lines, also a possible
asymmetry of the $V_{nn}^0$
distribution can be removed, as
announced below (\ref{180}).

Next we turn to the question of how far
perturbations which satisfy (\ref{260})
are ``weak'' in some physically meaningful 
sense.
Quite obviously, such considerations are
only possible in terms of non-rigorous
arguments and rough estimates.

First of all, typicality of thermalization, as invoked 
below (\ref{260}), trivially fails for vanishing 
perturbations
and hence may possibly still fail
for extremely weak perturbations
\cite{ale16,bra15}.
Yet, a closer inspection of the
non-perturbative approach from 
Refs.~\cite{deu91,rei15b}
suggests \cite{f4} that 
typicality of thermalization
generally 
does apply
provided
$\Gamma\gg D$ (cf. Eq.~(\ref{240})) 
and thus
\begin{eqnarray}
\sigma_v\gg D
\ .
\label{270}
\end{eqnarray}
In particular, the diagonal matrix elements
$V^0_{nn}$ are then typically much larger 
than the level spacing $D$,
thus corroborating the claim below (\ref{260}) 
that the ensembles of $\tilde V$'s and of $V$'s 
are often quite similar.

Second, while the unperturbed system $\tilde H_0$
is assumed not to thermalize for the given
initial condition $\rho(0)$,
one still expects that it exhibits the usual
thermodynamic properties when the system 
state happens to be the microcanonical
ensemble $\rhomic:=N^{-1}\sum_{n=1}^N
| n\rangle_{\! 0\, 0}\!\langle n|$ corresponding
to the energy window $\Imc :=[E-\delta E, E]$
introduced below Eq.~(\ref{160}).
Denoting by $\Omega (E)$ the number 
of energy levels $\tilde E_n^0$ 
below $E$,
by $k_B$ Boltzmann's constant, 
and by $S(E):=k_B\ln[\Omega(E)]$
the entropy, the temperature 
is thus given by $T(E):=1/S'(E)$.
Moreover, $\delta E$ must not exceed
$k_BT(E)$, otherwise the level density
would no longer be (approximately) 
constant throughout $\Imc $
(as assumed below Eq.~(\ref{160})).
It now seems reasonable to say that a perturbation
is weak if it does not notably change the
thermal equilibrium properties
($S(E)$, $T(E)$, heat 
capacity, state of matter, etc.)
of the unperturbed system.
Closer inspection of the approach
from Refs.~\cite{deu91,rei15b}
implies that the perturbations
are weak in this sense as long as
$\Gamma\ll k_BT(E)$
(otherwise, regions with different
level densities start to ``interact'' 
via the perturbation).
According to (\ref{240}) this amounts 
to $\sigma_v^2\ll k_BT(E)\, D$.
Focusing on the special (largest possible) 
choice $\delta E=k_BT(E)$ and 
exploiting $D:=\delta E/N$,
we arrive at $\sigma_v\ll \sqrt{N}D$ and
$\sigma_v^2\ll\delta E\, D$.
The first relation complements the
lower bound 
from (\ref{270}).
Since $N$ is exponentially 
large in the degrees of freedom,
the range of admitted $\sigma_v$ 
values is thus still very large.
The second relation agrees with
(\ref{260}) if $\trel$ is 
comparable to $\hbar/\delta E$.
As shown in Ref.~\cite{bal17}, this is 
indeed the case for a quite large class
of Hamiltonians $\tilde H_0$,
observables $A$, and initial 
conditions $\rho(0)$.

Alternatively, a perturbation
may be considered as weak
if the perturbed expectation value
$\Aobs(t)$ 
remains for (almost) all sufficiently 
large times $t$ close to the expectation value 
$\tr\{\rhomic A\}$,
which the unperturbed system {\em would\,}
assume in thermal equilibrium.
By similar arguments as above, one
can see that this alternative weak perturbation
criterion is 
essentially
equivalent to the one from the 
previous paragraph
{\em and\,} the condition (\ref{270}).

Altogether, Eq.~(\ref{260}) thus
seems to be a physically very 
natural weak perturbation condition, 
and it appears reasonable to conjecture that
prethermalization will
in general be ruled out 
if (\ref{260}) is violated.
We plan to further pursue this issue
in our future work.

In summary, prethermalization has been 
established for a very large class of integrable
(non-thermalizing) 
Hamiltonians $\tilde H_0$ 
and weak perturbations $\tilde V$, which 
closely imitate the essential features of 
many particular examples of interest in this context.
Adopting the common lore of random matrix 
theory \cite{fyo96,bor16,bro81,deu91},
the same conclusion is thus expected to
apply ``typically'' or ``with overwhelming likelihood'' also 
to any given such example,
unless there is some 
{\em a priori} reason 
(inappropriate choice of the ensemble,
another non-thermalizing system very 
near-by etc.)
why the specific 
example at hand must be one 
of the very rare exceptions 
with respect to {\em every} 
admitted $\tilde V$ ensemble
\cite{gol10a,gol10b,rei15b,rei15,bal17}.
Remarkably, the same predictions also 
apply to any other $\tilde H_0$ which exhibits 
equilibration but not thermalization, for 
instance due to many-body localization 
effects \cite{gog16,nan15,gol17}.

\begin{acknowledgments}
This work was supported by the 
Deutsche Forschungsgemeinschaft (DFG)
under Grant No. RE 1344/10-1 and
within the Research Unit FOR 2692
under Grant No. 397303734.
\end{acknowledgments}


\clearpage
\newpage

\onecolumngrid
\begin{center}
{\bf\large SUPPLEMENTAL MATERIAL}
\\[1cm]
\end{center}
\twocolumngrid
\setcounter{equation}{0}

Throughout this Supplemental Material, equations
from the main paper are indicated by an extra letter 
``m''. For example, ``Eq.~(m1)'' refers to 
Equation (1) in the main paper.

Sec. \ref{s1} provides the derivation of 
Eqs.~(m9)-(m13).

Sec. \ref{s2}  illustrates how a concrete physical
model system fits into our present
random matrix approach.

Sec. \ref{s3} numerically exemplifies
that the analytical results adopted in the 
main paper remain valid under considerably 
more general conditions.

Sec. \ref{s4} deduces Eqs.~(m15)-(m18) from 
the three assumptions above Eqs.~(m15), (m17), and (m18).

Sec. \ref{s5} provides a more detailed discussion of 
the statistical properties of  the random matrix elements 
$V^0_{mn}$.

Sec. \ref{s6} provides the derivation of Eq.~(m25).

\section{Derivation of Eqs.~($\mbox{m}$9)-($\mbox{m}$13)}
\label{s1}
In this section, Eqs.~(m9)-(m13)
will be derived.
Moreover, it will be shown that $Y(t)\in[0,1]$
and $1-Y(t)+W(t)\geq 0$, hence both roots
on the right hand side of (m10) are 
non-negative real numbers.

To begin with, we recall the definitions in Eqs. 
(m1)-(m8). We also recall that
the average over the random perturbations 
$V$ in Eq.~(m3) is indicated by an overline.
Accordingly,
$H_0$, $\rho(0)$, and $A$ are considered 
as arbitrary but fixed (non-random), 
hence the same follows for
$E_n^0$, $|n\rangle_{\! 0}$, $\rho_0(t)$,
$\rho^0_{mn}(t)$, and $A_{nm}^0$.
In contrast,
$H$, $E_n$, $|n\rangle$, $U_{mn}$, 
and $\gamma_{\tau\mu}(t)$ 
are random quantities.

Next, we rewrite (m7) as
\begin{eqnarray}
\Delta & = &  \alpha + \beta
\ , 
\label{a10}
\\
\alpha & := & 
\sum_{\mu\nu\sigma\tau} \rho_{\mu\nu}A_{\sigma\tau}\, 
[
\gamma_{\tau\mu} \gamma^\ast_{\sigma\nu} 
- 
\bar\gamma_{\tau\mu} \bar\gamma^\ast_{\sigma\nu}
]
\ ,
\label{a20}
\\
\beta & := & 
\sum_{\mu\nu\sigma\tau} \rho_{\mu\nu}A_{\sigma\tau}\, 
[
\bar\gamma_{\tau\mu} \bar\gamma^\ast_{\sigma\nu}
-
\delta_{\tau\mu}\delta_{\sigma\nu}
]
\ ,
\label{a30}
\end{eqnarray}
where, for notational convenience, all
arguments ``$t$'' and indices ``$0$'' are 
temporarily omitted.
Note that all factors on the right
hand side of (\ref{a30}) and hence 
$\beta$ on the left hand side
are non-random quantities.
In contrast, $\alpha$ in (\ref{a20}) is a random
quantity due to the random $\gamma$'s
on the right hand side.
Moreover, its average $\bar\alpha$
gives rise to a correlation-type quantity
on the right hand side of (\ref{a20}).
Roughly speaking, the main idea behind
(\ref{a10}) is to later split things
into such a correlation-type part 
and a ``rest''.

Denoting by $\amax$ and $\amin$
the largest and smallest eigenvalues of $A$,
and considering that the possible outcomes 
of a measurement are the eigenvalues
of the observable $A$, it follows that
\begin{eqnarray}
\da := \amax-\amin
\label{a40}
\end{eqnarray}
represents the measurement range of $A$.
Since all possible outcomes of any realistic 
measurement must be finite, we can take for 
granted that $\amax$, $\amin$, and $\da$ are finite.
Our next observation is that we can add an
arbitrary constant to $A$ without changing
the main quantity of interest, namely
the difference $\Delta (t)$ in (m6).
Without loss of generality we thus can
and will assume that $\amin=-\amax$, 
and hence
\begin{eqnarray}
\norm{A} = \da/2
\ ,
\label{a50}
\end{eqnarray}
where $\norm{A}$ denotes the operator
norm of $A$ (largest eigenvalue in modulus).
While $\Delta$ on the left hand side of (\ref{a10})
thus remains unchanged when adding some
constant to $A$, the same does not apply
separately to the two terms on the
right hand side, as can be seen
by closer inspection of (\ref{a20}) 
and (\ref{a30}).
Hence, while our special choice in (\ref{a50})
is irrelevant with respect to $\Delta$, it
does matter (and actually is optimized)
with respect to the splitting of $\Delta$
into $\alpha$ and $\beta$.

We recall that the $\rho_{\mu\nu}$ in (\ref{a20})
are -- according to Eq.~(m2) -- 
the matrix elements of $\rho_0(0)=\rho(0)$ 
in the unperturbed basis $|n\rangle_{\! 0}$, 
and likewise for $A_{\sigma\tau}$.
In the same vein, $\gamma_{\tau\mu}$
in (m8) may be viewed as the matrix 
elements 
$_0\!\langle \tau | \gamma | \mu \rangle_{\! 0}$
of an operator $\gamma$
(put differently, $\gamma$ is uniquely 
defined via those matrix elements).
Note that -- in view of (m8) --
this operator $\gamma$  may in 
general be non-Hermitian.
Likewise, the Kronecker delta
$\delta_{\tau\mu}$ in (\ref{a30})
may be viewed as the matrix 
elements of the identity operator.
As a consequence, one can rewrite
(\ref{a20}) and (\ref{a30}) as
\begin{eqnarray}
\alpha & = & 
\tr\{
\rho\gamma^\dagger A \gamma
\}
-
\tr\{
\rho\bar\gamma^\dagger A \bar\gamma
\}
\ ,
\label{a60}
\\
\beta & = & 
\tr\{
\rho\bar\gamma^\dagger A \bar\gamma
\}
-
\tr\{
\rho A 
\}
\ .
\label{a70}
\end{eqnarray}

%
\subsection{Evaluation of $\alpha$}
\label{s11}
In order to further evaluate $\alpha$ in (\ref{a60}), we define
the auxiliary (non-Hermitian, random) operator
\begin{eqnarray}
\eta := \gamma -\bar\gamma
\ .
\label{a80}
\end{eqnarray}
Replacing $\gamma$ in (\ref{a60}) by $\eta+\bar\gamma$,
a straightforward calculation yields
\begin{eqnarray}
\alpha 
& = & 
\alpha_1+\alpha_2+\alpha_2^\ast
\ ,
\label{a90}
\\
\alpha_1 
& := & 
\tr\{ \rho\eta^\dagger A\eta\}
\ ,
\label{a100}
\\
\alpha_2
& := & 
\tr\{ \rho\bar\gamma^\dagger A\eta\}
\ .
\label{a110}
\end{eqnarray}

By rewriting the right hand side of (\ref{a100}) as
$\tr\{A(\eta \rho\eta^\dagger)\}$,
observing that $\eta \rho\eta^\dagger$ is
a non-negative Hermitian operator,
and evaluating the trace by means of 
the eigenbasis of $A$, one finds that
$|\alpha_1|\leq \norm{A}\tr\{\eta \rho\eta^\dagger\}$.
With (\ref{a50}) we thus arrive at
\begin{eqnarray}
|\alpha_1| 
& \leq &
(\da/2)\, X
\ ,
\label{a120}
\end{eqnarray}
where
\begin{eqnarray}
X 
& := & 
\tr\{\eta \rho\eta^\dagger\}
\label{a130}
\end{eqnarray}
is a real-valued, non-negative random variable.

Turning to $\alpha_2$ in (\ref{a110}),
we note that $\rho$ is a non-negative 
Hermitian operator, hence
there exists a Hermitian operator, 
which we denote as $\sqrt{\rho}$, with
the property that $(\sqrt{\rho})^2=\rho$.
Exploiting the cyclic invariance of the
trace, the right hand side of 
(\ref{a110}) can thus be rewritten as
$\tr\{(\sqrt{\rho}\bar\gamma^\dagger A)(\eta\sqrt{\rho})\}$.
Considering $\tr\{B^\dagger C\}$
as a scalar product between two 
(not necessarily Hermitian) 
operators $B$ and $C$, the Cauchy-Schwarz
inequality takes the form
$|\tr\{B^\dagger C\}|^2\leq
\tr\{B^\dagger B\}\tr\{C^\dagger C\}$.
Choosing $B=A\bar\gamma\sqrt{\rho}$ 
and $C=\eta\sqrt{\rho}$ 
we can thus infer from (\ref{a110}) that
\begin{eqnarray}
|\alpha_2|^2\leq 
\tr\{\sqrt{\rho}\bar\gamma^\dagger A A \bar\gamma \sqrt{\rho}\}
\tr\{\sqrt{\rho}\eta^\dagger \eta \sqrt{\rho}\}
\ .
\label{a140}
\end{eqnarray}
Exploiting the cyclic invariance of the trace once more,
the second trace in (\ref{a140}) can be identified with
$X$ in (\ref{a130}).
Likewise, the first trace can be identified with
$\tr\{A^2 \bar\gamma\rho\bar\gamma^\dagger\}$.
Similarly as above (\ref{a120}), the latter trace
can be upper bounded by
$\norm{A}^2 Y$, where
\begin{eqnarray}
Y
:=
\tr\{\bar\gamma\rho\bar\gamma^\dagger\}
\ .
\label{a150}
\end{eqnarray}
Together with (\ref{a50}), we thus can conclude that
\begin{eqnarray}
|\alpha_2|^2\leq 
(\da/2)^2 \,X\, Y
\ .
\label{a160}
\end{eqnarray}

Next, we introduce $\eta$ from (\ref{a80}) into
(\ref{a130}) to obtain
\begin{eqnarray}
X & = & X_1 
- \tr\{\gamma\rho\bar\gamma^\dagger\}
- \tr\{\bar\gamma\rho\gamma^\dagger\}
+ \tr\{\bar\gamma\rho \bar\gamma^\dagger\}
\ ,
\label{a170}
\\
X_1 & := & \tr\{\gamma\rho\gamma^\dagger\}
=
\tr\{\rho\gamma^\dagger\gamma\}
=
\sum_{\mu\nu\sigma}
\rho_{\mu\nu}\gamma^\ast_{\sigma\nu}\gamma_{\sigma\mu}
\ .
\label{a180}
\end{eqnarray}
The last step follows along similar lines as 
above (\ref{a60}).
It follows that
\begin{eqnarray}
X_1 & = & \sum_{\mu\nu}
\rho_{\mu\nu} Q_{\mu\nu}
\ ,
\label{a190}
\\
Q_{\mu\nu} & := & 
\sum_{\sigma} \gamma_{\sigma\mu}\gamma^\ast_{\sigma\nu}
\ .
\label{a200}
\end{eqnarray}
From (m8), we can infer that
\begin{eqnarray}
\gamma_{\sigma\mu}\gamma^\ast_{\sigma\nu} =
& &  
\sum_{m} 
U^\ast_{m\sigma}U_{m\mu}\,
e^{i(E_\mu^0-E_m)t/\hbar}
\nonumber
\\
& & 
\times \sum_{n} 
U_{n\sigma}U^\ast_{n\nu}\,
e^{-i(E_\nu^0-E_n)t/\hbar}
\ ,
\label{a210}
\end{eqnarray}
hence (\ref{a200}) can be rewritten as
\begin{eqnarray}
Q_{\mu\nu} & = & \sum_{mn}
e^{i(E_\mu^0-E_m-E_\nu^0+E_n)t/\hbar}
U_{m\mu}U^\ast_{n\nu}\theta_{mn}
\ ,
\label{a220}
\\
\theta_{mn} & := & \sum_{\sigma}
U^\ast_{m\sigma}U_{n\sigma}
\ .
\label{a230}
\end{eqnarray}
Taking into account the definition (m4),
it follows that $\theta_{mn}=\delta_{mn}$,
hence (\ref{a220}) takes the form
\begin{eqnarray}
Q_{\mu\nu}=
e^{i(E_\mu^0-E_\nu^0)t/\hbar}
\sum_{n} U_{n\mu}U^\ast_{n\nu}
\ .
\label{a240}
\end{eqnarray}
Again, (m4) implies that the last sum
equals $\delta_{\mu\nu}$, hence 
$Q_{\mu\nu}=\delta_{\mu\nu}$, and 
(\ref{a190}) yields
\begin{eqnarray}
X_1=\sum_\nu\rho_{\nu\nu}=\tr\{\rho\}=1
\ .
\label{a250}
\end{eqnarray}

Technically speaking, the possibility
to exactly evaluate the products of four
$U$-matrix elements appearing in (\ref{a190})
via (\ref{a210}) is one of the key
points of our present approach.
In particular, this is the only place 
where such products (without averaging over
the $V$ ensemble) 
actually appear. 

Introducing (\ref{a250}) into (\ref{a170})
and averaging on both sides implies
\begin{eqnarray}
\bar X = 1 - \tr\{\bar\gamma\rho \bar\gamma^\dagger\}
\ .
\label{a260}
\end{eqnarray}
Upon comparison with (\ref{a150}), we can conclude
that
\begin{eqnarray}
Y=1-\bar X
\label{a270}
\end{eqnarray}
Since both $X$ in (\ref{a130}) and $Y$
in (\ref{a150}) must be non-negative real 
numbers, it follows that
\begin{eqnarray}
\bar X,\, Y \in[0,1]
\ .
\label{a280}
\end{eqnarray}

Observing that (\ref{a90}) implies
$|\alpha|\leq|\alpha_1|+2|\alpha_2|$,
we thus can infer from (\ref{a120}),
(\ref{a160}), and (\ref{a280}) that
\begin{eqnarray}
|\alpha|\leq (\da/2)\,(X+2\sqrt{X})
\ .
\label{a290}
\end{eqnarray}

%
\subsection{Evaluation of $\beta$}
\label{s12}
By means of the definition
\begin{eqnarray}
R := 
\tr\{\rho\bar\gamma^\dagger A\bar\gamma\}
+
\tr\{\rho A\bar\gamma\}
-
\tr\{\rho\bar\gamma^\dagger A\}
-
\tr\{\rho A\}
\label{a300}
\end{eqnarray}
one readily verifies that
\begin{eqnarray}
R^\ast = 
\tr\{\rho\bar\gamma^\dagger A\bar\gamma\}
+
\tr\{\rho\bar\gamma^\dagger A\}
-
\tr\{\rho A\bar\gamma\}
-
\tr\{\rho A\}
\label{a310}
\end{eqnarray}
and with (\ref{a70}) that
\begin{eqnarray}
\beta = (R+R^\ast)/2
\ .
\label{a320}
\end{eqnarray}
Rewriting (\ref{a300}) as
\begin{eqnarray}
R = 
\tr\{\rho(\bar\gamma^\dagger+1) A (\bar\gamma-1)\}
\label{a330}
\end{eqnarray}
it follows, similarly as above (\ref{a140}), that
\begin{eqnarray}
R = 
\tr\{[\sqrt{\rho}(\bar\gamma^\dagger+1) A]\,[ (\bar\gamma-1)\sqrt{\rho}]\}
\label{a340}
\end{eqnarray}
and hence
\begin{eqnarray}
|R|^2 & \leq &
\tr\{\sqrt{\rho}(\bar\gamma^\dagger+1) A^2 (\bar\gamma+1)\sqrt{\rho}\}\, S_{-}
\ ,
\label{a350}
\\
S_\pm & := & \tr\{\sqrt{\rho}(\bar\gamma^\dagger\pm 1)(\bar\gamma\pm 1)\sqrt{\rho}\}
\nonumber
\\
& = & \tr\{(\bar\gamma\pm 1) \rho (\bar\gamma^\dagger\pm 1)\}
\ .
\label{a360}
\end{eqnarray}
Similarly as above (\ref{a150}), the first factor on the
right hand side of (\ref{a350}) can be upper bounded by
$\norm{A}^2S_+$, yielding with (\ref{a50}) the result
\begin{eqnarray}
|R|^2\leq (\da/2)^2 S_+ S_{-}
\ .
\label{a370}
\end{eqnarray}
From (\ref{a320}) we can conclude that $|\beta|\leq |R|$ and hence
\begin{eqnarray}
|\beta|\leq (\da/2) \sqrt{S_+ S_{-}}
\ .
\label{a380}
\end{eqnarray}

Next, we rewrite (\ref{a360}) as
\begin{eqnarray}
S_\pm 
& = & 
\tr\{\bar\gamma \rho \bar\gamma^\dagger\}
\pm
\tr\{\rho \bar\gamma^\dagger\}
\pm
\tr\{\bar\gamma \rho\}
+
\tr\{\rho\}
\nonumber
\\
& = & Y \pm (Z^\ast + Z) +1
\ ,
\label{a390}
\\
Z & := & \tr\{\bar\gamma \rho\}
\ .
\label{a400}
\end{eqnarray}
In the second step in (\ref{a390}), we exploited 
$Z^\ast=\tr\{\rho \bar\gamma^\dagger\}$, $\tr\{\rho\}=1$, 
and Eq.~(\ref{a150}).
It follows that $S_+S_-=(1+Y)^2-(Z+Z^\ast)^2=(1-Y)^2+W$, where
\begin{eqnarray}
W:=4Y-(Z+Z^\ast)^2
\ ,
\label{a410}
\end{eqnarray}
and hence with (\ref{a270}) and (\ref{a380}) that
\begin{eqnarray}
|\beta|\leq (\da/2)\sqrt{\bar X^2+W}
\ .
\label{a420}
\end{eqnarray}
We finally remark that $S_+ S_- = \bar X^2+W$ 
is a real and non-negative number according 
to (\ref{a360}).
Due to (\ref{a280}) it follows that also
$W$ is a real number and satisfies 
$\bar X^2 + W\geq 0$.

%
\subsection{Evaluation of $\overline{|\Delta|}$}
\label{s13}
Eq.~(\ref{a10}) implies $|\Delta| \leq |\alpha|+|\beta|$
and hence $\overline{|\Delta|} \leq 
\overline{|\alpha|}+\overline{|\beta|}$.
Recalling that $\beta$ is a non-random quantity (see
below (\ref{a30})) yields
$\overline{|\beta|}=|\beta|$.
With (\ref{a290}) and (\ref{a420}) it then 
follows that
\begin{eqnarray}
\overline{|\Delta|} & \leq & \frac{\da}{2} \left(\bar X
+ 2\overline{\sqrt{X}}
+\sqrt{\bar X^2+W}\right)
\ .
\label{a430}
\end{eqnarray}
Note that since $X$ is a real-valued, 
non-negative random variable 
(see below (\ref{a130})), 
the same applies to $\sqrt{X}$.
Observing that $F(x):=\sqrt{x}$ is a concave 
function for all $x>0$, we can exploit Jensen's 
inequality to obtain
$\overline{\sqrt{X}}\leq \sqrt{\bar X}$.
Moreover, we can conclude from (\ref{a280}) 
that $\bar X\leq\sqrt{\bar X}$.
Altogether, (\ref{a430}) thus implies
\begin{eqnarray}
\overline{|\Delta|} & \leq & \frac{\da}{2} \left(3\sqrt{\bar X}+\sqrt{\bar X+W}\right)
\ .
\label{a440}
\end{eqnarray}

Finally, we can recast $Y$ from (\ref{a150})
by means of similar arguments as above 
(\ref{a60}) into the form
\begin{eqnarray}
Y=\tr\{\rho\bar\gamma^\dagger\bar\gamma\}
=\sum_{\mu\nu\sigma} \rho_{\mu\nu}
\bar\gamma^\ast_{\sigma\nu}\bar\gamma_{\sigma\mu}
\ . 
\label{a450}
\end{eqnarray}
Likewise, $Z$ from (\ref{a400}) takes the form
\begin{eqnarray}
Z=\tr\{\rho\bar\gamma\}
=\sum_{\mu\nu} \rho_{\mu\nu}
\bar\gamma_{\nu\mu}
\ . 
\label{a460}
\end{eqnarray}

Reintroducing the omitted indices ``$0$'' 
and arguments ``$t$''
(see below (\ref{a30})), 
we recover from 
(\ref{a270}),
(\ref{a400}),
(\ref{a440})-(\ref{a460})
our final result
\begin{eqnarray}
\overline{|\Delta(t)|} 
& \leq & 
\frac{\da}{2}\, f(t)
\ ,
\label{a465}
\\
f(t)
& := &
3\sqrt{1-Y(t)}+\sqrt{1-Y(t)+W(t)}
\ ,
\label{a470}
\\
Y(t)
& := & 
\sum_{\mu\nu\sigma} 
\rho^0_{\mu\nu}(t)\,
\overline{\gamma_{\sigma\mu}(t)}\,
\left[\overline{\gamma_{\sigma\nu}(t)}\right]^\ast
\ , 
\label{a480}
\\
W(t) 
& := & 
4Y(t) -[Z(t)+Z^\ast(t)]^2
\ ,
\label{a490}
\\[0.2cm]
Z(t)
& := & 
\sum_{\mu\nu} 
\rho^0_{\mu\nu}(t)\,
\overline{\gamma_{\nu\mu}(t)}
\ .
\label{a500}
\end{eqnarray}
These results are identical to
Eqs.~(m9)-(m13).
We also recall that $Y(t)$ and $W(t)$ are
known to be real numbers with $Y(t)\in[0,1]$
(see (\ref{a280})) and $1-Y(t)+W(t)\geq 0$
(see below (\ref{a420}) 
and above (\ref{a440})).
Altogether, we thus recover the findings 
announced at the beginning of this section.

\section{Random matrix description of a spin chain model}
\label{s2}
In this section, we illustrate the approximation 
of a concrete physical model 
system by the random matrix approach from the main paper.
As a particularly simple and common example
we consider an (integrable)
Heisenberg spin-1/2 chain in the presence of a weak 
integrability breaking perturbation.

Similarly as in the main paper, the perturbed Hamiltonian
is written in two alternative forms,
\begin{eqnarray}
H=H_0+V=\tilde H_0+ \tilde V \ ,
\label{b10}
\end{eqnarray}
where $\tilde H_0$ and $\tilde V$ are the
``bare'' 
operators,
while $H_0$ and $V$ are their 
``dressed'' counterparts.
The detailed definition of those 
operators will be provided below,
and the essential idea
behind those definitions will be
discussed in Sec. \ref{s23}.

\begin{figure*}
\includegraphics[scale=0.95]{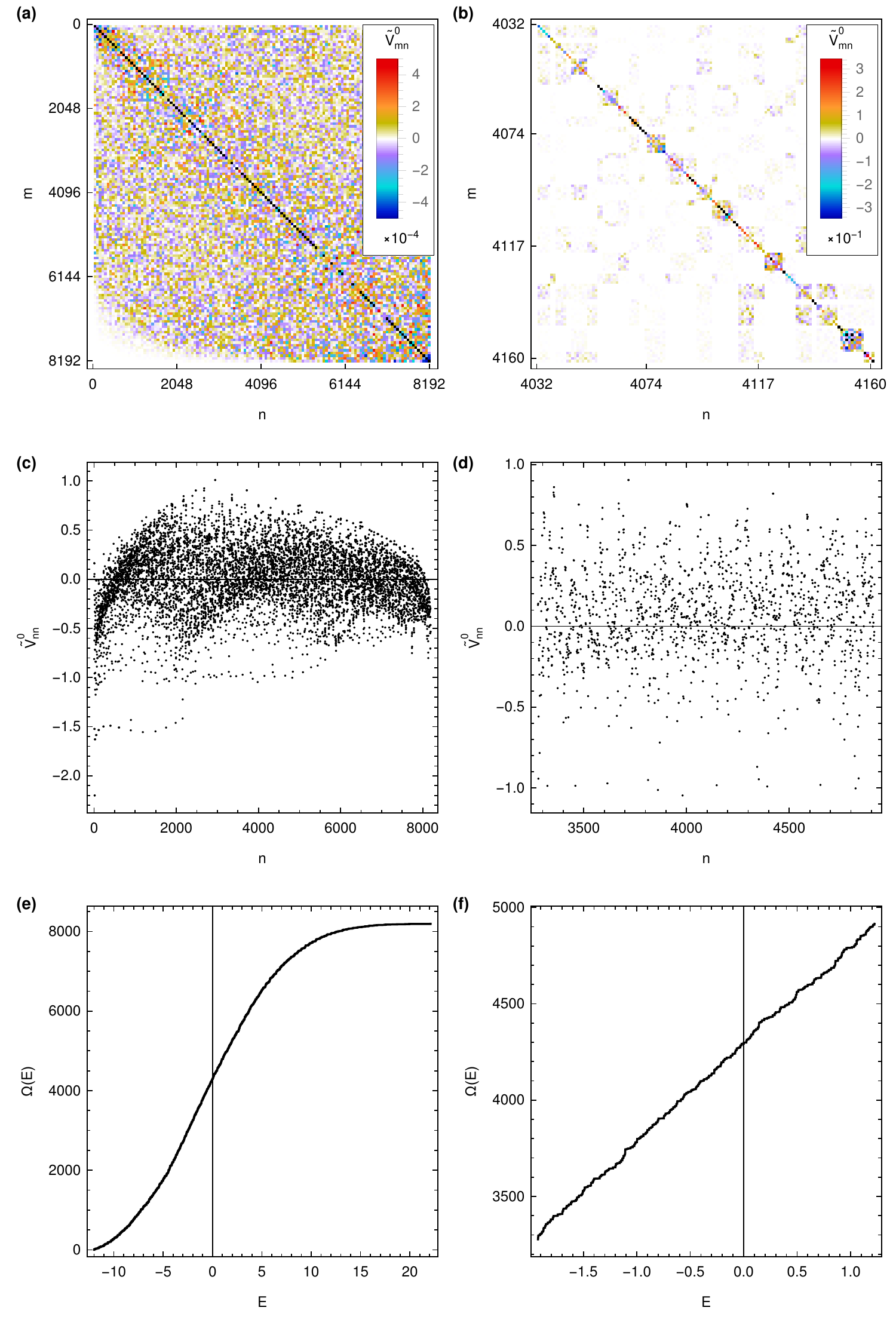}
\caption{Illustration of the
eigenvalues $\tilde E^0_n$ and the
matrix elements $\tilde V^0_{mn}$ for the Heisenberg 
spin-1/2 chain from Eqs.~\eqref{b20} and~\eqref{b30} 
with $L = 13$ and $\lambda = 0.2$.
Panels (b), (d), and (f) are close-up views of a central segment of 
Panels (a), (c), and (e), respectively.
In~(a) and~(b), we show the matrix elements $\tilde V^0_{mn}$ 
with their magnitude color-coded as indicated.
In~(a), one pixel corresponds to an average over blocks of 
$64 \times 64$ states.
In~(b), one pixel corresponds to one matrix element.
The scale is chosen to contrast off-diagonal matrix elements.
Since the diagonal matrix elements are on the average about 
two orders of magnitude larger than the off-diagonal ones, 
many of  them lie outside of the plotting range and are 
shown as black squares.
In~(c) and~(d), we show the diagonal elements $\tilde V^0_{nn}$ as a 
function of the index $n$.
In~(e) and~(f), the level counting function $\Omega(E)$ from 
Eq.~\eqref{b50} is plotted, illustrating the distribution of energy levels $\tilde E^0_n$.
In the middle of the spectrum shown in~(f), the level spacing is 
approximately constant, so that $\Omega(E)$ is a nearly straight line with 
slope $1/D$, where $D\simeq 0.002$ is the mean level spacing.
}
\label{fig1}
\end{figure*}

\subsection{The model}
\label{s21}
As announced, we focus on a 
perturbed Heisenberg spin-1/2 chain 
with ``bare'' operators (cf. Eq.~(\ref{b10}))
\begin{eqnarray}
\tilde H_0 & := & - J \sum_{l=1}^{L-1} 
\vec \sigma_l\cdot \vec\sigma_{l+1}
\ ,
\label{b20}
\\
\tilde V & := & - \lambda\sum_{l=1}^{L-2}\sigma^z_l\sigma_{l+2}^z
\ ,
\label{b30}
\end{eqnarray}
where 
$\sigma^{\alpha}_l$ (with $\alpha=x,y,z$ and $l=1,...,L$)
are Pauli matrices acting on site $l$
and $\vec\sigma_l:=(\sigma_l^x,\sigma_l^y,\sigma_l^z)$.
The unperturbed Hamiltonian (\ref{b20}) amounts to $L$
spins with nearest neighbor interactions and is well known to
be integrable, while the perturbation (\ref{b30}) consists of
integrability breaking next-to-nearest neighbor interactions.
The units are chosen so that
\begin{eqnarray}
J=\hbar=1
\ ,
\label{b40}
\end{eqnarray}
and the (in these units) dimensionless parameter $\lambda$
quantifies the perturbation strength.

Following the main paper,
the eigenvalues and eigenvectors of $\tilde H_0$ in (\ref{b20})
are denoted as $\tilde E^0_n$ and $|n\rangle_{\!0}$, 
and the matrix elements of the 
perturbation (\ref{b30}) are abbreviated as 
$\tilde V^0_{mn} :=\, _0\!\langle m | \tilde V | n \rangle_{\! 0}$.
For a spin chain of length $L=13$ and perturbation strength 
$\lambda = 0.2$, the unperturbed 
eigenvalues $\tilde E_n^0$ 
and the perturbation matrix 
$\tilde V^0_{mn}$
are illustrated in Fig.~\ref{fig1}.
In this example, the indices have been 
chosen so that the $\tilde E_n^0$'s are 
ordered by magnitude, where $n=1,...,2^L$ and
$2^L$ is the Hilbert space 
dimension of the model (\ref{b10}).
Closer inspections shows that it is always possible to 
choose the eigenvectors $|n \rangle_{\! 0}$ so that all 
matrix elements $\tilde V^0_{mn}$ are real numbers.
Without loss of generality, the eigenvectors have been 
chosen in this way in all numerical results of this section.

Fig.~\ref{fig1} depicts the so obtained
matrix
$\tilde V^0_{mn}$, 
the diagonal matrix elements
$\tilde V^0_{nn}$, and the 
energy levels $\tilde E^0_n$ in the first, 
second, and third rows, respectively.
In each case,
we show a view of the 
full Hilbert space in 
the left panels and a close-up of the 
middle of the spectrum in 
the right panels.

The coarse structure in Fig.~\ref{fig1}(a) already appears 
to be qualitatively ``random''.
Looking at the individual matrix elements in detail 
in Fig.~\ref{fig1}(b), 
however, reveals correlations that manifest themselves 
very roughly speaking as blocks
of vanishing and non-vanishing entries.
Comparing off-diagonal and diagonal matrix elements, we observe 
that the former are typically one to two orders of magnitude 
smaller than the latter.
For the rest, at least in the middle of the spectrum shown 
in Fig.~\ref{fig1}(d), the diagonal matrix 
elements essentially look like random numbers.

In Fig.~\ref{fig1}(e) and~(f), we display the distribution 
of energy levels $\tilde E^0_n$ by means of the function 
$\Omega (E)$ from the main paper,
which counts the number of $\tilde E_n^0$'s
with the property $\tilde E_n^0\leq E$.
Put differently,
\begin{eqnarray}
\Omega (E):=\sum_{n=1}^{2^L} \Theta(E-\tilde E_n^0)
\ ,
\label{b50}
\end{eqnarray}
where $\Theta (x):=\int_{-\infty}^x \delta(y)\, dy$ 
is the Heaviside step function.
The example in Fig.~\ref{fig1}(e)
nicely illustrates a property, which is 
generic for many-body systems, and which is 
taken for granted in the main paper,
namely that the energy spectrum
gives rise to a well-defined local level density, 
which may change notably only on scales much 
larger than the corresponding mean level 
spacing.
(Otherwise, the Boltzmann entropy 
$S(E) := k_{\mathrm B} \ln \Omega(E)$ would 
not lead to reasonable thermodynamic properties 
of the system, see end of the main paper.)
Here and in the following, we focus on the
middle of the spectrum, where the level density
is almost constant over a particularly large
energy interval, but similar results would be 
obtained for any other energy interval
which is not too large and not too close to 
the upper and lower ends of the spectrum.
(With increasing $L$, those restrictions are
expected to become weaker and weaker).

We numerically explored the above model along 
the same lines as 
for the dressed model in Sec. \ref{s22}
and found that it can {\em not} be satisfactorily 
approximated by our present random matrix 
approach.
The main reason seems to be
(see also Sec. \ref{s31} below)
that the diagonal matrix elements 
$\tilde V^0_{nn}$ in Fig.~\ref{fig1}(c)
exhibit rather large fluctuations (upon variation of $n$),  
which are not reflected in the considered random matrix 
model \cite{fyo96}.

\subsection{Dressed operators}
\label{s22}
The main purpose of the ``dressed'' operators $H_0$ 
and $V$ in (\ref{b10}) is to overcome the above 
mentioned mismatch between the bare operators
and our present random matrix approach.
To this end, we first combine the unperturbed levels 
$\tilde E^0_n$ with the diagonal matrix elements 
$\tilde V^0_{nn}$, defining
\begin{equation}
\label{b71}
	q_n := \tilde E^0_n + \tilde V^0_{nn} \,.
\end{equation}
Very roughly speaking, the idea behind this definition
and the following considerations is to
effectively absorb 
into the unperturbed Hamiltonian
those parts of the perturbation 
$\tilde V$ that do not break integrability, thus isolating the 
integrability-breaking non-diagonal contributions.
 
Given
that the fluctuations of
the $\tilde V^0_{nn}$'s are still much
smaller than the energy scale over which 
the level density exhibits notable variations, 
and that those fluctuations are essentially 
unbiased random numbers (see above),
one expects that also the $q_n$'s
in (\ref{b71}) exhibit some well-defined local density, 
which is moreover
practically equal to the local density of 
the $\tilde E_n^0$'s.
On the other hand, since the fluctuations of the 
$\tilde V^0_{nn}$ are significantly larger than 
the mean level spacing $D$, the so-defined 
$q_n$'s are no longer ordered by magnitude.
To restore this order, we relabel the 
eigenvectors $|n \rangle_{\! 0}$ so that 
$q_n \leq q_{n+1}$ for all $n$.
In all that follows, this modified
ordering of the labels $n$ is implicitly 
understood.
(Obviously, the physical properties of 
the problem at hand do not depend on 
how we order the $n$'s.)

\begin{figure*}
\includegraphics[scale=0.95]{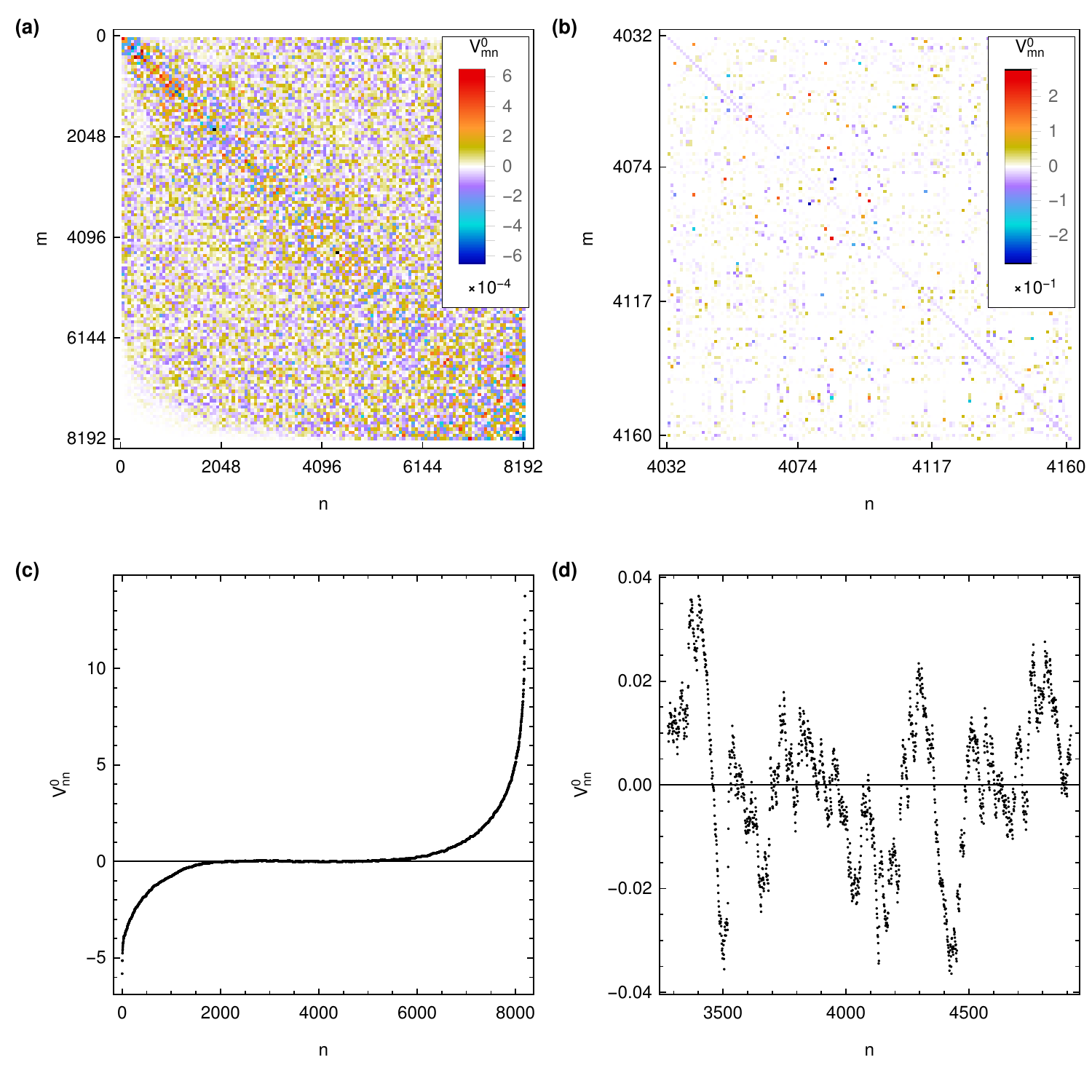}
\caption{Illustration of the dressed perturbation matrix elements 
$V^0_{mn}$ for the Heisenberg spin-1/2 chain from 
Eqs.~\eqref{b20} and~\eqref{b30} with $L = 13$ and 
$\lambda = 0.2$.
Panels~(b) and~(d) are close-up views of a central 
segment of Panels~(a) and~(c), respectively.
While the coarse structure of the off-diagonal elements 
in~(a) looks very similar to the bare case from 
Fig.~\ref{fig1}(a), 
the fine structure in~(c) now looks much more
``random'' than in Fig.~\ref{fig1}(b).
The dressed diagonal elements $V^0_{nn}$ 
are plotted in~(c) and~(d).
Compared to the bare diagonal elements from 
Fig.~\ref{fig1}(d), the dressed 
ones in~(d) now 
exhibit much smaller fluctuations.
In fact, the fluctuations of the diagonal elements
are now comparable to those of the
off-diagonal matrix elements.
}
\label{fig2}
\end{figure*}

As in the main paper, we finally require
that $H_0$ in (\ref{b10})
should exhibit the same eigenvectors 
$|n\rangle_{\!0}$ as $\tilde H_0$, whereas 
the eigenvalues $E_n^0$ must
be equally spaced, with a level spacing $D$ which
reproduces the {\em mean} level spacing of the 
$\tilde E_n^0$'s (or, equivalently, the $q_n$'s)
within the microcanonical energy 
window $\Imc :=[E-\delta E,E]$ of 
actual interest.
Recalling that $N$ denotes 
the (large but finite) number of energies
$\tilde E_n^0$ within $\Imc$, it follows 
that $D=\delta E/N$.
Accordingly, we set
\begin{equation}
\label{b91}
	E_n^0 := n D + c
\end{equation}
with a still arbitrary constant $c$.
In order to satisfy the second identity in Eq.~(\ref{b10}), 
the diagonal matrix elements of the dressed 
perturbation must thus be defined as
\begin{equation}
\label{b72}
	V^0_{nn} := q_n - E_n^0 \,,
\end{equation}
while the off-diagonal matrix elements must
remain unchanged, $V^0_{mn} := \tilde V^0_{mn}$,
apart from the above mentioned
reordering of the labels $n$.

\begin{figure*}
\includegraphics[scale=1]{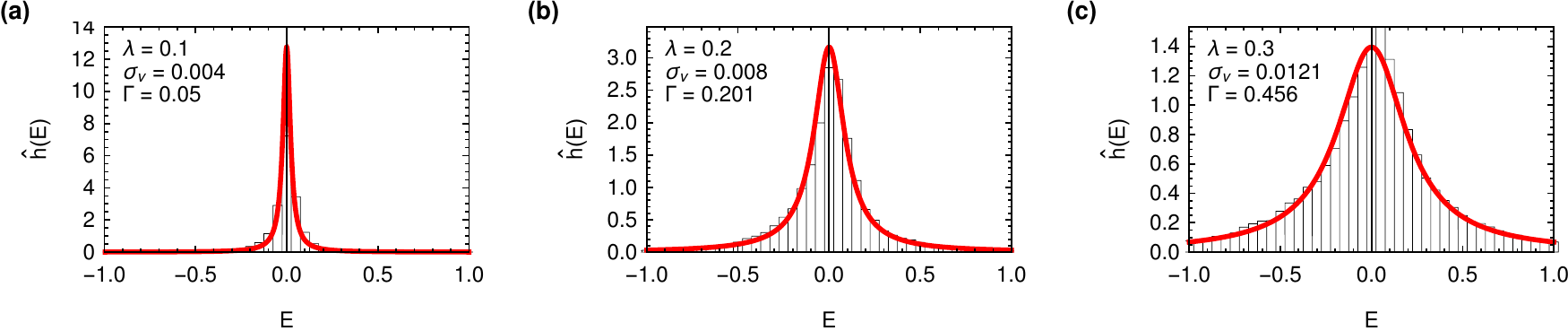}
\caption{Numerical local density of states~\eqref{b140} for the spin-1/2 Heisenberg model from 
Eqs.~\eqref{b20} and~\eqref{b30} with $L = 13$ for (a) $\lambda = 0.1$, (b) $\lambda = 0.2$, and 
(c) $\lambda = 0.3$.
The histograms show the binned numerical data as described in the text.
The solid lines correspond to the theoretical prediction~\eqref{b120}--\eqref{b130} with $\sigma_v^2$ 
and $D$ extracted from the selected energy shell, i.e., the central $20 \%$ of states.
The width of the pre-delta function in~\eqref{b150} is $\epsilon = 0.05$.
}
\label{fig3}
\end{figure*}

By way of their above construction,
we expect that the dressed operators
$H_0$ and $V$ now meet the 
characteristics of generic random matrix ensembles such 
as those from Ref.~\cite{fyo96}.
This is verified numerically in Fig.~\ref{fig2} for the 
same model and data as in Fig.~\ref{fig1}, 
but presented in 
terms of the dressed matrix $V$ instead of the bare
$\tilde V$.
(The distribution of energies $E^0_n$ as defined in Eq.~\eqref{b91} 
is rather boring to look at and therefore not plotted.)
As the microcanonical energy window, we selected the 
central $20\%$ of states, and the constant $c$ in 
Eq.~\eqref{b91} was chosen such that the average 
of the $V^0_{nn}$'s vanishes within this window.

While the coarse view in Fig.~\ref{fig2}(a) looks 
similar to the one from Fig.~\ref{fig1}(a),
the close-up in Fig.~\ref{fig2}(b) shows 
that also the fine structure of the perturbation 
now indeed looks very much like
a ``true'' (possibly sparse) random 
matrix.
The diagonal elements $V_{nn}^0$ are
separately reproduced in Fig.~\ref{fig2}(c) and~(d).
In particular, Fig.~\ref{fig2}(d) confirms that the local
fluctuations are now much 
smaller than those of 
$\tilde V_{nn}^0$ in Fig.~\ref{fig1}(e).

Similarly as when comparing ``true'' 
random numbers with numerically 
generated pseudo-random numbers,
it is in general very difficult to 
quantitatively determine how closely the 
$V_{mn}^0$'s in Fig.~\ref{fig2} 
emulate a ``true'' random 
matrix.
We therefore focus on the
specific quantity which is
at the heart of our present work, 
namely the (shifted) local density of 
states (see also (m21))
\begin{eqnarray}
h_n(E)
& := &
\sum_{m} |U_{mn}|^2\,\delta(E-E_m+E_n^0)
\ ,
\label{b100}
\\
U_{mn} & := & \langle m|n\rangle_{\!0}
\ ,
\label{b110}
\end{eqnarray}
where $E_m$ and $|m\rangle$
denote (as in the main paper) 
the eigenvalues and eigenvectors 
of the perturbed system $H$ in (\ref{b10}).

As detailed in the main paper, 
the ensemble average of $h_n(E)$ is 
theoretically predicted to be given by
\begin{eqnarray}
\bw (E) & := &
\frac{1}{2\pi} \frac{\Gamma}
{E^2+\Gamma^2/4}
\label{b120}
\ ,
\\
\Gamma & := & 2\pi \sigma_{\! v}^2/D \,,
\label{b130}
\end{eqnarray}
where $\sigma_v^2$ is the ensemble 
averaged variance of the $V^0_{mn}$'s
(with $m\not=n$ and $|m-n|$ not too large).
In our numerics, this prediction is 
compared with
\begin{eqnarray}
\hat h (E) := \frac{1}{N}\sum_{n\in I_N} h_n(E)
\ ,
\label{b140}
\end{eqnarray}
where $I_N$ represents the set of indices with
$E_n^0\in \Imc$ and $N$ is their number.
Furthermore, $\sigma_v^2$ in (\ref{b140})
is now the numerically determined
variance of the $V_{mn}^0$'s.
Finally, the delta function appearing in 
(\ref{b120}) is replaced by a pre-delta 
function 
\begin{eqnarray}
\delta_\epsilon(x):= 
\left\{
\begin{array}{ll}
1/\epsilon & \mbox{if $|x|\leq \epsilon/2$} \\
0 & \, \textrm{otherwise} \\
\end{array}
\right. ,
\label{b150}
\end{eqnarray}
essentially amounting to a binning of the
numerical data with bin width $\epsilon$.

Estimating the ensemble average of
$h_n(E)$ from (\ref{b100})
by the column average $\hat h(E)$ 
of a single realization in Eq.~\eqref{b140} is common practice in 
random matrix theory (sometimes referred 
to as self-averaging or ergodicity property),
and is justified by 
the large number of states $N$ 
within the microcanonical energy window and the 
translational invariance of the ensemble of perturbations, 
meaning that $V^0_{mn}$ and $V^0_{m+k, n+k}$ 
exhibit 
similar statistical properties for all $k$ 
and thus may be viewed as different
samples of the random matrix 
ensemble.
In particular, the column average in 
Eq.~\eqref{b140} is thus expected to converge
for large $N$ towards the ensemble
average of $h_n(E)$.

While the theoretical predictions strictly speaking apply 
to infinitely large matrices, the numerically
considered matrices are of large but
finite dimension $2^L = 8192$. In order to minimize
 finite size artifacts, 
which are known to be particularly pronounced 
near the borders of the matrices,
we choose the interval $I_N$ to comprise the central 
$20 \%$ of states as before, such that $N = 1638$.

In Fig.~\ref{fig3}, we compare the so 
obtained numerical results for the function 
$\hat h(E)$ with the theory from Eqs.~\eqref{b120} 
and~\eqref{b130} for three different perturbation 
strengths $\lambda$ in (\ref{b30}).
We emphasize that there are no fit parameters:
The width $\Gamma$ of the Breit-Wigner distribution 
\eqref{b120} is
obtained from the values of $\sigma_v^2$ (see figure)
and $D = 0.002$ 
computed from the actual matrix elements $V^0_{mn}$ and 
energy levels $\tilde E^0_n$.
The plots reveal good agreement between numerics and theory, 
even for the moderate perturbation strength $\lambda = 0.3$, 
where $\Gamma / \delta E \approx 0.14$.

\subsection{Concluding remarks}
\label{s23}
Finally, it may be worthwhile to recall
the following key ideas from the main paper
regarding the connection of the above
considerations with the main objectives
of the paper:
In general, the two alternative
unperturbed Hamiltonians $H_0$ and $\tilde H_0$
in (\ref{b10}) give rise to different time-dependent
expectation values.
However, the difference of those expectation values
can be bounded with the help of our inequality (m25).
Closer inspection of how the right hand sides
of (m24) and of (m25) numerically scale with
the number of spins $L$ in (\ref{b20})
indicates that for sufficiently large $L$ the
left hand side of (m25) in fact becomes 
arbitrarily small (compared to $\da$)
for all times $t$ with $|t|\leq 1/\Gamma$.
Since we are mainly interested in large $L$
and $|t|\leq 1/\Gamma$ in our present work,
the two unperturbed Hamiltonians
$H_0$ and $\tilde H_0$ thus give rise to
expectation values which can be
considered as practically indistinguishable.
Therefore it is justified to work with the
dressed perturbation $V$ in (\ref{b10})
instead of the bare $\tilde V$ in our
above considerations.

\begin{figure*}
\includegraphics[scale=1]{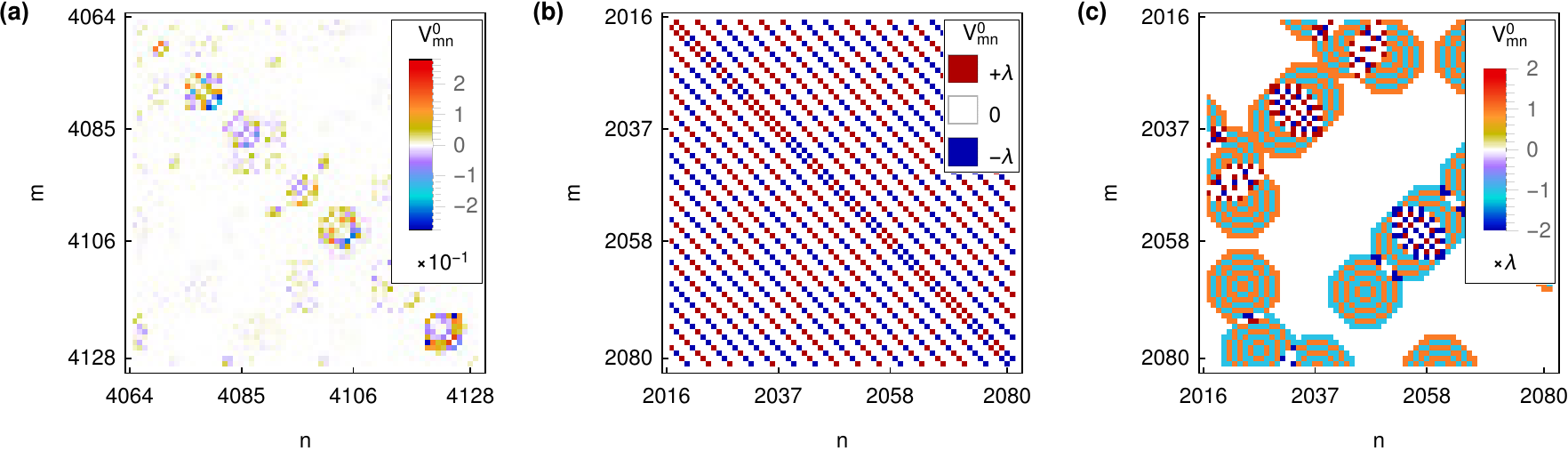}
\caption{Exemplary submatrix view of the three types of perturbations 
considered in Sec.~\ref{s3}:
(a) the spin model from Sec.~\ref{s2} for $\lambda = 0.2$, but with
the diagonal entries set to zero, cf.~Fig.~\ref{fig1}(a,b);
(b) a matrix with regular sparsity ($r = 4$) and nonvanishing entries 
randomly sampled from $\{-\lambda, +\lambda\}$;
(c) a matrix with random sparsity and regular radial patterns
(see \eqref{b180}, \eqref{b190})
for islands of nonvanishing entries with 
$s_{\mathrm{isl}} = 15$.
For further details see text.
}
\label{fig4}
\end{figure*}

Besides those times with $|t|\leq 1/\Gamma$,
also the long-time limit plays an essential
role.
Specifically, the long-time average of (m1)
takes the form
\begin{eqnarray}
\langle \Aobs_0 (t)\rangle_t=
\sum_{n} \rho^0_{nn}(0)\, A^0_{nn}
\label{b160}
\end{eqnarray}
where $\langle \Aobs_0 (t)\rangle_t$ indicates
the long-time average of $\Aobs_0 (t)$,
and where we exploited that the spectrum
of $H_0$ is non-degenerate (see (\ref{b91})).
In particular, this long-time
average refers to the dynamics governed
by $H_0$.
But since the eigenvectors of $\tilde H_0$ and
$H_0$ are identical, the above
result must also be equal to the long-time
average when the dynamics is governed by
$\tilde H_0$.
(In contrast to $H_0$, the spectrum of $\tilde H_0$
may exhibit degeneracies; in such a case,
the eigenvectors $|n\rangle_{\! 0}$ are tacitly
chosen so that $\rho(0)$ is diagonal within
every energy eigenspace.)
One thus can conclude that also with respect
to the long-time properties it is justified
to work with the dressed perturbation $V$
in (\ref{b10}) instead of the bare
$\tilde V$.

\section{Numerical illustration of generalized 
random matrix ensembles}
\label{s3}
In this section we provide three illustrative numerical
examples, indicating that the analytical predictions
from Ref.~\cite{fyo96} remain valid for considerably 
more general random matrix ensembles than those
actually admitted therein
(see also main paper below (m24)).
In particular, while in random matrix theory 
it is usually assumed that all matrix elements 
of some given ensemble are 
statistically independent of each other, 
we will demonstrate that many 
results, e.g.~Eqs.~\eqref{b120} and~\eqref{b130}, 
are still applicable in the presence of various,
quite notable correlations.
Since such correlations naturally arise in 
real systems (see also Sec.~\ref{s2}), this 
is an important observation with regard to their 
modeling by random matrices.

The general framework and the notation
is as in the previous section (and in the 
main paper).
In particular, we again compare the
analytical prediction (\ref{b120}) with the
numerically obtained counterpart (\ref{b140}),
i.e., we will numerically estimate the ensemble 
average of $h_n(E)$ from (\ref{b100})
by the column average $\hat h(E)$ 
of a single realization in Eq.~\eqref{b140}.
While the theory strictly speaking applies to
infinitely large matrices, our numerical examples 
will be of large but
finite dimension $M$. In order to minimize
the corresponding finite size artifacts, 
which are known to be particularly pronounced 
near the borders of the matrices,
the interval $I_N$ in (\ref{b140}) is chosen
so that it remains sufficiently far away 
from the lower and upper limits
at $n=1$ and $n=M$, respectively.
As before, the delta function, which enters
into (\ref{b140}) via (\ref{b100}), is approximated
by the pre-delta function (\ref{b150}),
effectively amounting to a binning of 
the numerical data.

\subsection{First example}
\label{s31}
As our first example, we again start out from 
the spin model \eqref{b20}--\eqref{b30} 
with $L=13$, hence $M = 2^L = 8192$.
But now, we modify the actual perturbation $\tilde V$ 
by manually setting all diagonal elements $\tilde V^0_{nn}$ 
equal to zero, or equivalently, we consider
dressed operators which no longer satisfy 
(\ref{b10}) but rather are defined via
\begin{equation}
\label{b170}
	H_0 := \tilde H_0 \,,
	\quad
	V^0_{mn} := \tilde V^0_{mn} - \delta_{mn} \tilde V^0_{nn} \,.
\end{equation}
While the resulting Hamiltonian $H_0 + V$ is no longer expected
to faithfully model some ``natural'' physical system, it still amounts 
to an instructive ``artificial'' example:
On the one hand, the large fluctuations of the diagonal elements
$\tilde V^0_{nn}$ (see Fig.~\ref{fig1}(c)) are eliminated.
On the other hand, the remaining off-diagonal elements
still exhibit a considerable amount of ``structures'' or 
``correlations'' (see Figs.~\ref{fig1}(b)
and \ref{fig4}(a)).
Such a perturbation matrix $V^0_{mn}$ may thus be 
considered either as an extremely ``unlikely/exceptional/untypical''
member of one of  the random matrix ensembles admitted in 
the analytical explorations from \cite{fyo96},
or as a ``typical'' member of an ensemble which is not 
admitted in those analytical explorations.

\begin{figure*}
\includegraphics[scale=1]{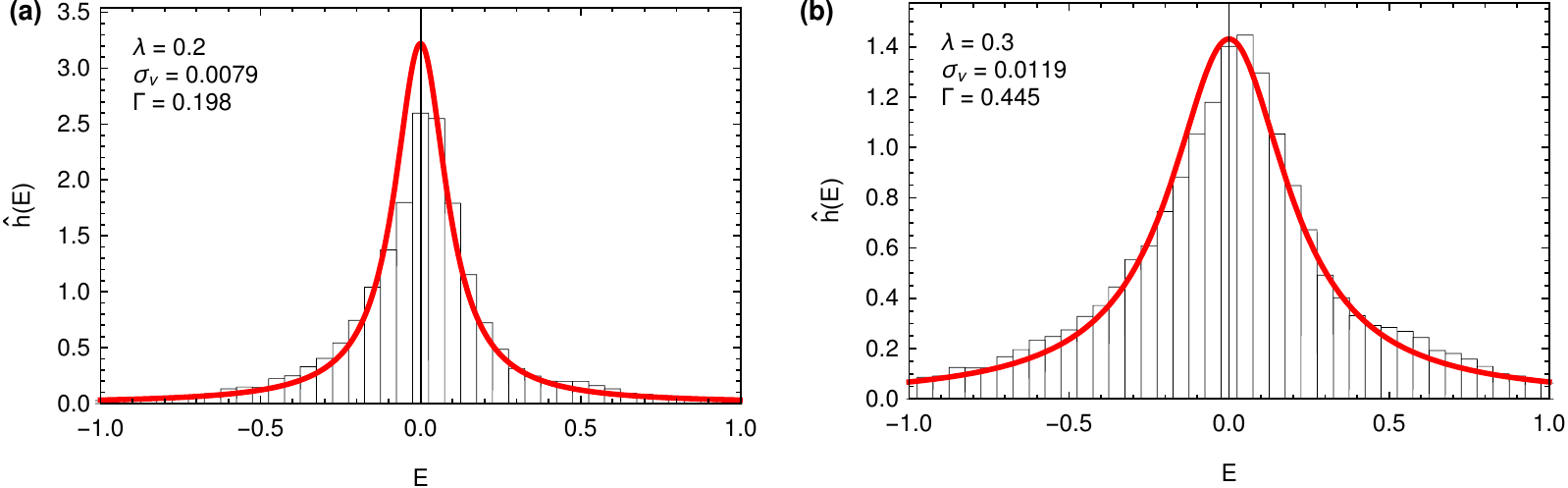}
\caption{Numerical local density of states~\eqref{b140} (histograms
with bin width $\epsilon = 0.05$)
and theoretical prediction \eqref{b120}--\eqref{b130} (red) for the
modified spin model (see also main text and Fig. \ref{fig4}(a))
with $L = 13$ and (a) $\lambda = 0.2$, (b) $\lambda = 0.3$.
}
\label{fig5}
\end{figure*}

\begin{figure*}
\includegraphics[scale=0.975]{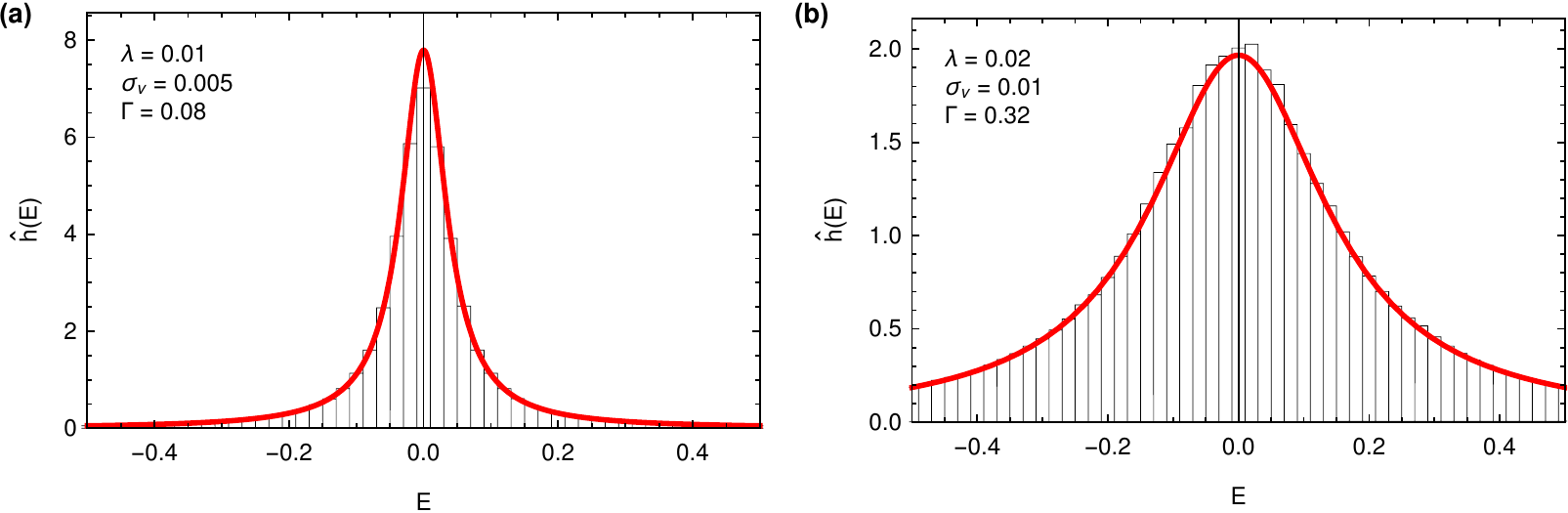}
\caption{Numerical local density of states~\eqref{b140} (histograms)
and theoretical prediction (\eqref{b120},\eqref{b130}) (red)
 for $H_0$ 
with Poissonian level statistics and a perturbation matrix with 
regular sparsity $r = 4$ and random entries $\pm\lambda$ 
with (a) $\lambda = 0.01$ and (b) $\lambda = 0.02$.
The solid lines correspond to the theory curve from 
Eqs.~\eqref{b120} and~\eqref{b130}.
The bin width is $\epsilon = 0.02$.
}
\label{fig6}
\end{figure*}

Focusing, as in the previous section, 
on a microcanonical energy window consisting of the central $20\%$ 
of states, we numerically evaluated $\hat h(E)$ from~\eqref{b140}, the 
local density of states $1/D$, as well as the sample variance 
of perturbation elements to estimate $\sigma_v^2$.
Hence there are again no free parameters in the theory.
Comparing $\hat h(E)$ to the Breit-Wigner formula~\eqref{b120} with 
width~\eqref{b130} in Fig.~\ref{fig5}, 
we observe that the theory still describes the 
numerical estimate quite well. 
In particular, this substantiates the assertion
at the end of Sec. \ref{s21}.

\begin{figure*}
\includegraphics[scale=1]{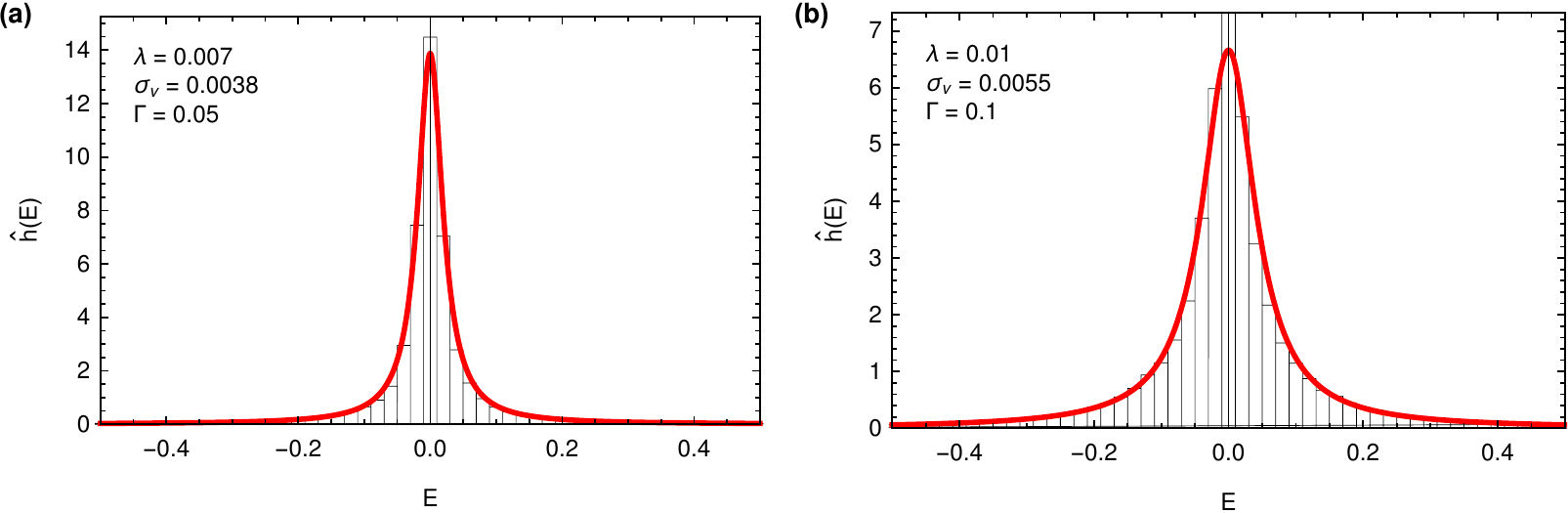}
\caption{Numerical local density of states~\eqref{b140} for 
$H_0$ with Poissonian level statistics and a perturbation matrix 
with $M_{\mathrm{isl}} = 16384$ randomly scattered islands 
of diameter $s_{\mathrm{isl}} = 15$ with entries given by
Eqs.~\eqref{b180}, \eqref{b190}
for (a) $\lambda = 0.007$ and (b) $\lambda = 0.01$.
The solid lines correspond to the theory curve from 
Eqs.~\eqref{b120} and~\eqref{b130}.
The bin width is $\epsilon = 0.02$.
}
\label{fig7}
\end{figure*}

\subsection{Further examples}
\label{s32}
In the following two examples, the total Hilbert space 
dimension is $M = 2^{12} = 4096$ and the unperturbed 
system is chosen to exhibit Poissonian level statistics 
with a mean spacing $D = 1/512 \approx 0.002$, reflecting 
the distribution of gaps of typical integrable systems.
(Equally spaced levels as in (\ref{b100}) or Wigner-Dyson 
distributed spacings would lead to practically the 
same results.)
As before,
the considered perturbations are quite artificial 
and are not meant to reflect actual physical systems.
Rather, their purpose is to illustrate the generality 
of the result~\eqref{b120}, \eqref{b130}.

In our first artificial ensemble of perturbations, 
only every $r$th entry starting from the first 
minor diagonal is nonvanishing.
Hence the resulting matrices are sparse with a 
very regular entry pattern.
Moreover, the nonvanishing entries can only take 
values $+\lambda$ and $-\lambda$, each with 
probability $\frac{1}{2}$.
An example of such a perturbation matrix is shown in 
Fig.~\ref{fig4}(b).
The resulting numerical estimate $\hat h(E)$ along 
with the theory prediction is shown in 
Fig.~\ref{fig6} for $r = 4$ 
and two different choices of $\lambda$.
As before, there are no fit parameters involved, 
since $D$ was fixed \emph{a priori} and 
$\sigma_v^2 = \lambda^2 / r$ by construction.
Even though the dimension of the considered 
matrices is rather small, at least vastly smaller than 
the size of the Hilbert space for typical many-body 
systems, theory and numerics agree very well.

For our second artificial example, we uniformly scatter $M_{\mathrm{isl}}$ 
circular islands of diameter $s_{\mathrm{isl}}$ across the 
perturbation matrix.
Within each island, we alternate entries $+\lambda$ and $-\lambda$ 
in the radial direction.
More precisely, if $(\hat m_k, \hat n_k)$ denotes the 
(randomly sampled) center of the $k$th island and 
$r^{(k)}_{mn} := \sqrt{ (m-\hat m_k)^2 + (n-\hat n_k)^2 }$ 
is the distance of entry $(m, n)$ from that center, then we 
define the contribution to $V^0_{mn}$ from the $k$th 
island as
\begin{equation}
\label{b180}
	v^{(k)}_{mn}
		:= \lambda \, (-1)^k \left( 2 \lfloor r^{(k)}_{mn} \;\mathrm{mod}\; 2 \rfloor - 1 \right) \, \Theta(s_\mathrm{isl}/2 - r^{(k)}_{mn}) \,,
\end{equation}
where `$\mathrm{mod}$' denotes the modulo operation and 
$\lfloor x \rfloor = q$ is the greatest integer $q$ such that 
$q \leq x$ (``floor function'').
Islands may overlap, in which case the contributions are 
simply added, i.e.,
\begin{eqnarray}
V^0_{mn} = V^0_{nm} := \sum_{k=1}^{M_{\mathrm{isl}}} v^{(k)}_{mn}
\ .
\label{b190}
\end{eqnarray}
Hence in this ensemble, the sparsity is irregular (random), but the 
distribution of nonvanishing entries shows a regular pattern.
For a visualization of a single realization of this ensemble, see Fig.~\ref{fig4}(c).

In Fig.~\ref{fig7}, we display the 
numerical estimate~\eqref{b140} of the local density 
of states for a perturbation with $M_{\mathrm{isl}} = 2^{14} = 16384$ 
islands of size $s_{\mathrm{isl}} = 15$ and for two different 
perturbation strengths $\lambda = 0.007$ and $\lambda = 0.01$.
The resulting average sparsity $\bar r = 4.1$ is similar as in 
the first artificial example, i.e., approximately every fourth matrix 
element $V^0_{mn}$ is nonvanishing.
The parameter $\sigma_v^2$ is computed as the sample 
variance of all perturbation matrix elements, so that as 
before there is no free parameter in the theory~\eqref{b120}.
Again, numerics and theory show good agreement.

We also explored several further examples of 
artificially ``structured'' or ``correlated'' perturbations
$V$ along the same lines as for
 those depicted in Fig. \ref{fig4};
for instance, patterns generated by 
cellular automata or by concentric 
rings (``corner arcs'') 
around the upper left corner 
of the matrix, etc.
We also investigated separately the 
impact of various ``patterns'' along the 
matrix diagonal.
Usually, the results were similar to those 
in Figs. \ref{fig5}-\ref{fig7} and are therefore 
not reproduced here.
Notably different results were only
obtained for examples with extremely
``little'' randomness or extremely
strong correlations, for instance, 
when choosing deterministically 
alternating signs along the off-diagonals in 
Fig. \ref{fig4}(b).
(In this case, the only random ingredient is
the Poissonian level statistics of the 
unperturbed spectrum.)

%
\section{Derivation of Eqs. ($\mbox{m}$15)-($\mbox{m}$18)}
\label{s4}
In this section, we deduce Eqs.~(m15)-(m18)
from the three assumptions 
above Eqs.~(m15), (m17), and (m18).

In doing so, our principal tool is
the Taylor expansion formula 
for functions of noncommuting operators 
derived in Ref.~\cite{kum65}.
It states that for two arbitrary linear
(but not necessarily Hermitian)
operators $A,B$ 
and any complex analytic function $\phi(z)$,
\begin{equation}
\label{d10}
	\phi(A+B) = 
         \sum_{n=0}^\infty \frac{1}{n!} C_n(A, B) \, \phi^{(n)}(A) 
         \,,
\end{equation}
where $\phi(A)$ is defined via the power series 
representation of $\phi(z)$, and
where $\phi^{(n)}(z)$ denotes the $n$th 
derivative of $\phi(z)$.
Moreover, the operators $C_n(A, B)$ are
given by the recurrence relation
\begin{equation}
\label{d20}
	C_0 := 1 \,, \quad
	C_n := \left[ A, C_{n-1} \right] + B C_{n-1}
\end{equation}
with the square brackets denoting the commutator.

A second key point is the observation that 
$\gamma_{\tau\mu}(t)$ from (m8) 
can be rewritten in terms of the 
function $\phi(z) := e^{-i z t / \hbar}$ 
as
\begin{equation}
\label{d30}
	\gamma_{\tau\mu}(t) 
         = e^{i E^0_\mu t / \hbar} \,_0\!\langle \tau | 
         \, \phi(H_0 + V) \, | \mu \rangle_{\! 0} 
         \ ,
\end{equation}
where $t$ is considered as arbitrary but fixed.
Using the Taylor expansion formula~\eqref{d10} 
with $A := H_0$ and $B:=V$, and noting that 
the above defined $\phi(z)$ is an analytical function
with $\phi^{(n)}(z) = (-i t / \hbar)^n \phi(z)$, 
we obtain
\begin{equation}
\label{d40}
	\gamma_{\tau\mu}(t) = 
        \sum_{n=0}^\infty \frac{(-i t / \hbar)^n}{n!} 
        \,_0\!\langle \tau | C_n | \mu \rangle_{\! 0}
\ .
\end{equation}
Furthermore, (\ref{d20}) now takes the form
\begin{equation}
\label{d50}
	C_0 := 1 \,, \quad C_n := [H_0, C_{n-1}] + V C_{n-1} \,.
\end{equation}
More explicitly, the first few $C_n$'s are
\begin{eqnarray}
	C_0 &=& 1 , \nonumber\\
	C_1 &=&V , \nonumber\\
	C_2 &=& [H_0, V] + V^2 , \nonumber\\
	C_3 &=& \left[ H_0, [H_0, V] \right] + [ H_0, V^2 ] 
        + V [ H_0, V ] + V^3 \,.\ \ 
\label{d60}
\end{eqnarray}
As exemplified by $C_2$ and $C_3$, the operators $C_n$
are thus not necessarily Hermitian.
Finally, the definitions (m8) and (m16) 
imply that $g_{\mu}(t)=\gamma_{\mu\mu}(t)$, 
and with (\ref{d40}) we thus obtain
\begin{equation}
\label{d70}
	\overline{ g_{\mu}(t) } = 
        \sum_{n=0}^\infty \frac{(-i t / \hbar)^n}{n!} 
        \overline{ \,_0\!\langle \mu | C_n | \mu \rangle_{\! 0} }
\ .
\end{equation}

\subsection{Derivation of Eqs. (m15) and (m16)}
\label{s41}
The goal of this subsection is to show 
that the assumption above Eq.~(m15)
implies
\begin{eqnarray}
\overline{\gamma_{\tau\mu}(t)}=0
\label{d80}
\end{eqnarray}
whenever $\mu\not=\tau$.
From this result and Eq.~(m8),
one readily recovers 
Eqs.~(m15) and (m16).

In order to verify (\ref{d80}), it is according 
to (\ref{d40}) sufficient to show that 
\begin{eqnarray}
\overline{\,_0\!\langle \tau | C_n | \mu \rangle_{\! 0} } = 0
\label{d90}
\end{eqnarray}
for all $\tau \neq \mu$ and any given $n\in\NN_0$.
To this end, we take for granted the 
assumption above Eq.~(m15).
In particular, the two unperturbed basis vectors
$|\tau\rangle_{\! 0}$, $|\mu\rangle_{\! 0}$
appearing in the average $\overline{\,_0\!\langle \tau | C_n | \mu \rangle_{\! 0} }$
may thus be multiplied by arbitrary 
factors $\sigma_{\tau},\sigma_{\mu}\in\{\pm 1\}$
without changing the value of that
average.

Due to the trivial fact 
that the operators 
$C_n$ in (\ref{d50}), (\ref{d60}) are 
independent of the basis vectors 
$|\tau\rangle_{\! 0}$, $|\mu\rangle_{\! 0}$ 
and thus of the factors 
$\sigma_{\tau},\sigma_{\mu}$,
it follows that
\begin{equation}
\label{d100}
	\overline{\,_0\!\langle \tau | C_n | \mu \rangle_{\! 0} }
	=  \sigma_\tau \sigma_\mu \,
           \overline{\,_0\!\langle \tau | C_n | \mu \rangle_{\! 0} }
\end{equation}
for arbitrary $\sigma_{\tau},\sigma_{\mu}\in\{\pm 1\}$.
If $\tau=\mu$, then $\sigma_\tau \sigma_\mu =\sigma_{\tau}^2=1$,
hence (\ref{d100}) is always fulfilled.
If $\tau\not=\mu$, the factor $\sigma_\tau \sigma_\mu$
may assume both values $\pm 1$,
hence (\ref{d100}) implies (\ref{d90}).

\subsection{Justification of Eq. (m17)}
\label{s42}
The objective of this subsection is to  show
that the function $\overline{g_\mu(t)}$ 
is independent of $\mu$ and hence $g(t)$ 
from Eq.~(m17) is well-defined,
provided the assumption above 
Eq.~(m17) is fulfilled.

The assumption above Eq.~(m17) 
consist of two parts:
The first part requires that 
$E_{n+1}^0-E_n^0=D$ for all $n$ and thus
\begin{eqnarray}
E_m^0-E_n^0=(m-n)\, D
\label{d110}
\end{eqnarray}
for all $E_m^0,E_n^0$.
The second part requires that 
all statistical properties of the matrix elements 
$V^0_{mn}$ do not depend separately on $m$ and $n$, 
but only on the difference $m-n$.
In particular, this implies that
\begin{equation}
\label{d120}
	\overline{ V_{\nu_0 \nu_1} V_{\nu_1 \nu_2} \cdots V_{\nu_{k-1} \nu_k} } = v_k \left( \{ \nu_i - \nu_{i-1}\}_{i=1}^k \right)
\end{equation}
for arbitrary indices $\nu_0,...,\nu_k$ and any $k\in\NN$, 
where the right hand side indicates that
the left hand side is given by some function 
$v_k$ which only depends on the differences
$\nu_1-\nu_0, \nu_2-\nu_1,..., \nu_k-\nu_{k-1}$.

In view of Eq.~(\ref{d70}), the objective 
from the beginning of this subsection
will be achieved if we can infer from 
(\ref{d110}) and (\ref{d120}) that 
$\overline{ \,_0\!\langle \mu | C_n | \mu \rangle_{\! 0} }$ 
is independent of $\mu$ for all $n$.
The latter will be shown in what follows.

From~\eqref{d50} or~\eqref{d60} it can be seen that 
the $C_n$ are composed of powers of $V$ and commutators 
of these powers with $H_0$.
Working in the eigenbasis of $H_0$ and inserting complete sets 
of states between factors of $V$, we find that any diagonal 
matrix element $\,_0\!\langle \mu | C_n | \mu \rangle_{\! 0}$ 
can be written as a sum of terms of the general form
\begin{eqnarray}
&&
F(\mu, k, s_1, \ldots, s_{k}) 
\nonumber
\\
&&
= \sum_{\nu_1, \ldots, \nu_{k-1}} V^0_{\mu \nu_1} (E^0_{\nu_1} - E^0_{\mu})^{s_1} V^0_{\nu_1 \nu_2} (E^0_{\nu_2} - E^0_{\nu_1})^{s_2} 
\nonumber
\\
&&
\qquad\qquad\qquad \cdots V^0_{\nu_{k-1} \mu} (E^0_{\mu} - E^0_{\nu_{k-1}})^{s_{k}}
\label{d130}
\end{eqnarray}
with $s_i \in \mathbb{N}_0$ and with the property 
that 
\begin{eqnarray}
k + \sum_{i=1}^k s_i = n
\ .
\label{d131}
\end{eqnarray}
The explicit expression of 
$\,_0\!\langle \mu | C_n | \mu \rangle_{\! 0}$ 
in terms of the functions (\ref{d130})
which satisfy (\ref{d131})
is given in Ref.~\cite{kum65}, 
but does not matter for the 
following arguments.

From Eqs.~\eqref{d120} and (\ref{d130})
we can conclude that
\begin{eqnarray}
&&\!\!\!\!\!\!
\overline{ F(\mu, k, s_1, \ldots, s_{k}) }
\nonumber
\\
&&\!\!\!\!\!\!
= \sum_{\nu_1, \ldots, \nu_{k-1}} (E^0_{\nu_1} - E^0_{\mu})^{s_1} (E^0_{\nu_2} - E^0_{\nu_1})^{s_2} 
           \cdots (E^0_{\mu} - E^0_{\nu_{k-1}})^{s_{k}} 
\nonumber
\\
&&\!\!\!\!\!\!
\qquad\qquad\quad \times v_k( \nu_1 - \mu, \{ \nu_i - \nu_{i-1} \}_{i=2}^{k-1}, \mu - \nu_{k-1} ) \,.
\label{d140}
\end{eqnarray}
As usual, the value of the sum on the right hand 
side does not change when a summation 
index $\nu_i$ is shifted by an arbitrary 
integer.
In particular, we thus may substitute
each index $\nu_i$ by 
$\nu_i+\mu$ ($i=1,...,k-1$).
Exploiting (\ref{d110}) it follows that
\begin{eqnarray}
&&\!\!\!\!\!\!
\overline{ F(\mu, k, s_1, \ldots, s_{k}) }
\nonumber
\\
&&\!\!\!\!\!\!
= \sum_{\nu_1, \ldots, \nu_{k-1}} (E^0_{\nu_1})^{s_1} (E^0_{\nu_2} - E^0_{\nu_1})^{s_2} \cdots (- E^0_{\nu_{k-1}})^{s_{k}}
\nonumber
\\
&&\!\!\!\!\!\!
\qquad\qquad\quad \times  v_k(\nu_1, \{ \nu_i - \nu_{i-1} \}_{i=2}^{k-1}, -\nu_{k-1}) \,,
\label{d150}
\end{eqnarray}
where the right hand side is manifestly independent of $\mu$.
Since all $\overline{ \,_0\!\langle \mu | C_n | \mu \rangle_{\! 0} }$  
are sums of such terms (see above), it follows that
they must be independent of $\mu$.

\subsection{Derivation of Eq.~(m18)}
\label{s43}
The goal of this subsection is to show 
that the assumptions above 
Eqs.~(m15), (m17), and (m18)
imply
\begin{eqnarray}
\overline{ g_\mu(t) } = \overline{ [g_\mu(t)]^\ast }
\ .
\label{d160}
\end{eqnarray}
Since also (m17) is known to apply
under those assumptions, 
Eq.~(m18) then 
readily follows.

In passing we note that
(m16) implies 
$g_\mu(-t)=[g_\mu(t)]^\ast$, 
and with Eq.~(m17) 
it follows that $g(-t)=[g(t)]^\ast$.
Together with (\ref{d160}), 
we thus obtain
\begin{eqnarray}
g(t)= g(-t)
\ ,
\label{d170}
\end{eqnarray}
which is a quite interesting result in itself.

In order to verify (\ref{d160}), it is according 
to (\ref{d70}) sufficient
to show that the averaged diagonal matrix elements
$\overline{ \,_0\!\langle \mu | C_n | \mu \rangle_{\! 0} }$ 
are purely real for even $n$ and purely imaginary for odd $n$.
To this end, we recall that 
$\,_0\!\langle \mu | C_n | \mu \rangle_{\! 0}$
is given by a sum of terms of the form
(\ref{d130}).
Hence, in order to verify (\ref{d160}), it
is sufficient to show that
\begin{equation}
\label{d180}
	\overline{ F(\mu, k, s_1, \ldots, s_k) }
		= (-1)^n  \left( \overline{ F(\mu, k, s_1, \ldots, s_k) } \right)^*
\, ,
\end{equation}
where $n$ is determined by $k$ and 
$s_1, \ldots, s_k$ according to (\ref{d131}).

Taking for granted the assumption 
above (m17), it follows, as 
in the previous subsection, that 
Eqs.~(\ref{d110}), (\ref{d120}),
and (\ref{d150}) apply.

Without loss of generality we can and
will choose the energy scale so that 
$E_0^0=0$ and hence $E_{-m}^0=-E_m^0$
according to (\ref{d110}).
Together with (\ref{d131})
we thus can conclude that
\begin{eqnarray}
& & 
\!\!\!\!\!\!\!\!\!
(E^0_{-\nu_1})^{s_1} (E^0_{-\nu_2} - E^0_{-\nu_1})^{s_2} 
\cdots (-E^0_{-\nu_{k-1}})^{s_k}
\nonumber
\\
& & 
\!\!\!\!\!\!\!\!\!
=
(-1)^{n-k} \, 
(E^0_{\nu_1})^{s_1} (E^0_{\nu_2} - E^0_{\nu_1})^{s_2} \cdots (-E^0_{\nu_{k-1}})^{s_k}
\, .\ \
\label{d190}
\end{eqnarray}

Furthermore, the assumptions above~(m15) 
and~(m18) guarantee that the statistical 
properties of $-V^0_{mn}$ are identical to 
those of $V^0_{mn}$ for all $m, n$, 
implying that the functions
$v_k$ from~\eqref{d120} satisfy
\begin{equation}
\label{d200}
	v_k(\{ \nu_i - \nu_{i-1}\}_{i=1}^k ) 
        = (-1)^k \, v_k(\{ \nu_i - \nu_{i-1} \}_{i=1}^k) 
\,.
\end{equation}
Since $V$ is Hermitian, $V^0_{mn} = (V^0_{nm})^*$, 
it follows from (\ref{d120}) that
\begin{equation}
\label{d201}
v_k(\{ \nu_i - \nu_{i-1} \}_{i=1}^k)^\ast
=
v_k(\{ \nu_{i-1} - \nu_{i} \}_{i=1}^k)
\end{equation}
and with~\eqref{d200} that
\begin{equation}
\label{d210}
	v_k(\{ \nu_i - \nu_{i-1} \}_{i=1}^k ) 
        = (-1)^k \, v_k(\{ \nu_{i-1} - \nu_{i} \}_{i=1}^k )^* 
\,.
\end{equation}

Similarly as below (\ref{d140}) one sees that
all summation indices $\nu_i$ on the right
hand side of Eq.~\eqref{d150} may be replaced
by $-\nu_i$, yielding
\begin{eqnarray}
&&\!\!\!\!\!\!
\overline{ F(\mu, k, s_1, \ldots, s_{k}) }
\nonumber
\\
&&\!\!\!\!\!\!
= \sum_{\nu_1, \ldots, \nu_{k-1}} (E^0_{-\nu_1})^{s_1} (E^0_{-\nu_2} - E^0_{-\nu_1})^{s_2} \cdots (- E^0_{-\nu_{k-1}})^{s_{k}}
\nonumber
\\
&&\!\!\!\!\!\!
\qquad\qquad\quad \times  v_k(-\nu_1, \{ \nu_{i-1} - \nu_{i} \}_{i=2}^{k-1}, \nu_{k-1}) \,.
\label{d220}
\end{eqnarray}
Upon introducing \eqref{d190} and exploiting 
\eqref{d210} with $\nu_0=\nu_k=0$,
it follows that
\begin{eqnarray}
&&\!\!\!\!\!\!
\overline{ F(\mu, k, s_1, \ldots, s_{k}) }
\nonumber
\\
&&\!\!\!\!\!\!
= (-1)^n \!\!\! \sum_{\nu_1, \ldots, \nu_{k-1}} \!\!\! (E^0_{\nu_1})^{s_1} (E^0_{\nu_2} - E^0_{\nu_1})^{s_2} \cdots (- E^0_{\nu_{k-1}})^{s_{k}}
\nonumber
\\
&&\!\!\!\!\!\!
\qquad\qquad\qquad \times  v_k(\nu_1, \{ \nu_{i} - \nu_{i-1} \}_{i=2}^{k-1}, -\nu_{k-1})^* \,.
\label{d230}
\end{eqnarray}
Together with~\eqref{d150} we thus 
recover Eq.~\eqref{d180}.

%
\section{Statistics of the random matrix elements $V^0_{mn}$}
\label{s5}
In this section, the statistical properties of 
the the random matrix elements 
$V^0_{mn} :=\, _0\!\langle m | V | n \rangle_{\! 0}$
are discussed in more detail, especially those
mentioned in the main text below Eq.~(m23).

To begin with, we observe that 
the random distribution of 
any given matrix element $V_{mn}^0$ is
captured by the probability density
\begin{eqnarray}
p(v,m,n):=\overline{\delta(V_{mn}^0 - v)}
\ .
\label{e40}
\end{eqnarray}
Here, it is understood that
for random matrix ensembles with purely
real elements $V_{mn}^0$, only
real arguments $v$ are admitted in (\ref{e40}).
On the other hand, for ensembles 
with complex $V_{mn}^0$ 
(which is only possible for $m\not =n$),
the arguments $v$ in (\ref{e40})
are understood to be complex
for $m\not =n$ and real for $m=n$.
We also recall that
\begin{eqnarray}
\delta(z):=\delta({\rm Re}(z))\, \delta({\rm Im}(z))
\label{e50}
\end{eqnarray}
for complex arguments $z:=V_{mn}^0 - v$
on the right hand side of (\ref{e40}).
It follows that
\begin{eqnarray}
\delta(-z) & = & \delta(z)
\label{e60}
\end{eqnarray}
both for real and complex 
arguments $z$.

Next, the following two conditions
\begin{eqnarray}
p(-v,m,n) & = & p(v,m,n)
\ ,
\label{e10}
\\
p(v,m,n) & = & p(v,m-n,0)
\ ,
\label{e20}
\end{eqnarray}
will be deduced from the three 
conditions above Eqs.~(m15), (m17), 
and (m18).
These results (\ref{e10}), (\ref{e20})
readily imply the statements
below (m23), namely that
the statistics of any given matrix element
 $V_{mn}^0$ only depends
on $m-n$, and that $V_{mn}^0$ and 
$-V_{mn}^0$ are equally likely.

The assumption above Eq.~(m15)
(see also Sec. \ref{s41}) 
together with the remarks 
below (\ref{e40}) readily imply 
\begin{eqnarray}
\overline{\delta(\sigma_m\sigma_n V_{mn}^0 - v)}
=
\overline{\delta(V_{mn}^0 - v)}
\label{e80}
\end{eqnarray}
for arbitrary
$\sigma_{m},\sigma_{n} \in\{\pm 1\}$.
If $m=n$, 
then $\sigma_{m}\sigma_{n}=\sigma_m^2=1$,
hence (\ref{e80}) is always fulfilled.
If $m\not = n$,  the factor $\sigma_m\sigma_n$ 
in (\ref{e80}) may assume both values 
$\pm 1$. With (\ref{e40}) and (\ref{e60})
it thus follows that (\ref{e80})
is equivalent to
\begin{eqnarray}
p(-v,m,n)=p(v,m,n) 
\label{e90}
\end{eqnarray}
for all $m>n$.
(Unlike in (\ref{e10}), the case 
$m=n$ is still missing.)

Similarly, from the condition above Eq. 
(m17)
(see also Sec. \ref{s42})
one can readily deduce
(\ref{e20}) for all $m\not= n$.

Finally, 
the condition above Eq.~(m18) 
(see also Sec. \ref{s43})
in combination with
(\ref{e40}) and (\ref{e60})
implies
\begin{eqnarray}
p(-v,n,n)=p(v,n,n)
\label{e100}
\end{eqnarray}
for all $n$.
Hence, (\ref{e100}) together with (\ref{e90})
yields (\ref{e10}).

In conclusion,
the three conditions on $V^0_{mn}$
above Eqs.  (m15), (m17), 
and (m18)
imply the two conditions
(\ref{e10}) and (\ref{e20}) on $p(v,m,n)$.

From now on we focus, as in the main text, 
on the simplest and most common case 
that all $V_{mn}^0$ with $m\geq n$  
are statistically independent of each 
other. 
More precisely, since 
$V_{nm}^0=(V_{mn}^0)^\ast$,
we tacitly assume that $m\geq n$
in the above and the following statements.

As a consequence, all statistical
properties of the random matrix
are (by definition) fully captured
by the distributions of the single elements 
$V^0_{mn}$ in (\ref{e40}).
If those distributions satisfy 
(\ref{e10}) and (\ref{e20}),
it is straightforward to invert the above
line of reasoning with the final 
conclusion that the two conditions 
(\ref{e10}) and (\ref{e20}) imply 
the three conditions above 
Eqs.~(m15), (m17), and (m18).

In other words,
under the additional assumption 
that the matrix elements 
$V^0_{mn}$ are statistically independent 
for all $m\geq n$, 
the conditions (\ref{e10}) and (\ref{e20}) are actually
equivalent to the three conditions above 
Eqs.~(m15), (m17), and (m18).

%
\section{Derivation of Eq. ($\mbox{m}$25)}
\label{s6}
Throughout this section, we adopt the
same notation as in the main paper,
except that all indices 
``$0$'' are omitted for 
notational convenience.

Accordingly, we consider
two arbitrary density 
operators $\rho(t)$ and $\tilde\rho(t)$
with identical initial conditions,
\begin{eqnarray}
\rho(0)=\tilde\rho(0) \ ,
\label{f10}
\end{eqnarray}
but whose time evolution is
governed by two different 
Hamiltonians, namely
\begin{eqnarray}
H:=\sum_n E_n\,|n\rangle\langle n|
\ ,
\label{f20}
\\
\tilde H:=\sum_n \tilde E_n\,|n\rangle\langle n|
\ .
\label{f30}
\end{eqnarray}
In other words, the eigenvectors
$|n\rangle$ of $H$ and $\tilde H$
must be identical, while the
eigenvalues $E_n$ and 
$\tilde E_n$ may be different.

In case the dynamics is governed by $H$,
the expectation value of any given 
observable $A$ at time $t$ can be 
written, similarly as in Eq.~(m1),
in the form
\begin{eqnarray}
\Aobs(t) & = & 
\tr\{\rho(t)A\}
\ ,
\label{f40}
\\
\rho(t) 
& = & 
\propa_t \rho(0)\, \propa_t^\dagger
\ ,
\label{f50}
\\
\propa_t 
& := &
e^{-iHt/\hh} \ .
\label{f60}
\end{eqnarray}
The right hand side of (\ref{f60}) 
is understood as usual:
\begin{eqnarray}
e^{-iHt/\hh} := \sum_n e^{-iE_nt/\hh}\,|n\rangle\langle n|
\ .
\label{f70}
\end{eqnarray}

Likewise, if $\tilde H$ governs the dynamics,
the expectation value of $A$ at time $t$ 
can be written as 
\begin{eqnarray}
\tilde \Aobs(t)
& = & 
\tr\{\tilde\rho(t)A\}
\ ,
\label{f80}
\\
\tilde\rho(t) 
& = & 
\tilde \propa_t \rho(0) \, \tilde\propa_t^\dagger
\ ,
\label{f90}
\\
\tilde \propa_t 
& := & 
e^{-i\tilde Ht/\hh} = \propa_t' \, \propa_t
\ ,
\label{f100}
\\
\propa_t' 
& := &
e^{i(H-\tilde H)t/\hh} \ .
\label{f110}
\end{eqnarray}
The last identity in (\ref{f100})
relies on the fact that $H$ from (\ref{f20})
and $\tilde H$ from (\ref{f30}) commute.
Together with (\ref{f40})-(\ref{f70}) 
it follows that $\tilde\rho(t)=\propa_t'\rho(t)(\propa_t')^\dagger$
and due to the cyclic invariance of the trace that
\begin{eqnarray}
\delta 
& := &  
\Aobs(t)- \tilde \Aobs(t)
=
\tr\{\rho(t)\,B_t\}
\ ,
\label{f120}
\\
B_t & := &
A-(\propa_t')^\dagger A \, \propa_t'\ ,
\label{f130}
\end{eqnarray}
where the dependence of $\delta$
on $t$ has been omitted for the 
sake of simplicity.

The main goal of this section 
is to show that
\begin{eqnarray}
|\delta|
\leq \da \, |t| \, 
\max\limits_{n}|\tilde E_n-E_n|/\hbar
\label{f140}
\end{eqnarray}
for arbitrary $t$ and $A$,
where $\da$ is the difference 
between the largest and smallest 
eigenvalues of $A$.
Eq.~(m25)
then follows upon taking into account:
(i) As mentioned at the beginning of this 
section, all missing indices ``$0$'' have
to be restored.
(ii) As said below Eq.~(m16), 
only $n\in\{1,...,N\}$
actually count in (\ref{f140}).

In order to verify (\ref{f140}),
we first evaluate the trace in 
(\ref{f120}) by means of the  
eigenbasis of $\rho(t)$, 
yielding 
\begin{eqnarray}
|\delta | & \leq & \max_{\norm{\psi}=1} |\delta_{\psi}|
\ ,
\label{f150}
\\
\delta_\psi & := & \langle\psi|B_t|\psi\rangle \ ,
\label{f160}
\end{eqnarray}
where the maximization in (\ref{f150}) is
over all normalized vectors $|\psi\rangle$.

For an arbitrary but fixed 
vector $|\psi\rangle$ of unit norm
we can rewrite (\ref{f160}) with (\ref{f130}) as
\begin{eqnarray}
\delta_\psi 
& = & 
\langle\psi|A|\psi\rangle -
\langle\psi'|A|\psi'\rangle
\ ,
\label{f170}
\\
|\psi'\rangle 
& := & \propa_t'|\psi\rangle \ .
\label{f180}
\end{eqnarray}
With the definition 
\begin{eqnarray}
|\chi\rangle := |\psi'\rangle-|\psi\rangle
\label{f190}
\end{eqnarray}
we can conclude that
\begin{eqnarray}
\langle\psi'|A|\psi'\rangle
& = &
\langle\psi'|A|\psi\rangle + d_1
\ ,
\label{f200}
\\
d_1 
& := &
\langle\psi'|A|\chi\rangle
\ ,
\label{f210}
\\
\langle\psi'|A|\psi\rangle
& = &
\langle\psi|A|\psi\rangle + d_2
\ ,
\label{f220}
\\
d_2
& := &
\langle\chi|A|\psi\rangle \ .
\label{f230}
\end{eqnarray}
Eqs.~(\ref{f170}), (\ref{f200}), and (\ref{f220}) 
imply
\begin{eqnarray}
|\delta_\psi| \leq |d_1| + |d_2| \ .
\label{f240}
\end{eqnarray}
From the definition (\ref{f210}) and the
Cauchy-Schwarz inequality it follows that
\begin{eqnarray}
|d_1|^2=|\langle\chi| (A|\psi'\rangle)|^2
\leq
\langle\chi|\chi\rangle \langle\psi'|A^2|\psi'\rangle \ .
\label{f250}
\end{eqnarray}
Since we assumed that $|\psi\rangle$ is normalized,
also $|\psi'\rangle$ in (\ref{f180})
will be normalized and the last factor 
in (\ref{f250}) can be upper bounded by
$\norm{A^2}=\norm{A}^2$,
where $\norm{A}$ is the operator norm of $A$
(largest eigenvalue in modulus).
Exactly the same upper bound can be
obtained for $d_2$ in (\ref{f230}).
With (\ref{f240}) we thus arrive at
\begin{eqnarray}
|\delta_\psi| \leq 2 \norm{A}\sqrt{\langle\chi|\chi\rangle}
\ .
\label{f260}
\end{eqnarray}
Obviously, $\delta$ in (\ref{f120}) remains 
unchanged when adding an arbitrary real constant 
$c$ to $A$.
Hence, the inequality (\ref{f260}) with $\norm{A+c}$
instead of $\norm{A}$ on the right 
hand side remains valid for 
arbitrary $c$.
The minimum over all $c$ is assumed
when the largest and smallest eigenvalues of
$A+c$ are of opposite sign and equal modulus,
yielding
\begin{eqnarray}
|\delta_\psi| \leq  \da \sqrt{\langle\chi|\chi\rangle} \ ,
\label{f270}
\end{eqnarray}
where $\da$ is the difference
between the largest and smallest 
eigenvalues of $A$.

Rewriting $|\psi\rangle$ as $\sum_n c_n\,|n\rangle$
with $c_n:=\langle n|\psi\rangle$,
the normalization of $|\psi\rangle$
takes the form $\sum_n |c_n|^2=1$.
Furthermore, we can infer from
(\ref{f20}), (\ref{f30}), (\ref{f110}), 
and (\ref{f180}) that
\begin{eqnarray}
|\psi'\rangle & = & \sum_n e^{i a_n} c_n |n\rangle
\ ,
\label{f280}
\\
a_n & := & (E_n-\tilde E_n)t/\hh
\label{f290}
\end{eqnarray}
and from (\ref{f190}) that
\begin{eqnarray}
\langle\chi|\chi\rangle
=
\sum_n |c_n-e^{ia_n}c_n|^2
=
\sum_n |c_n|^2 \,|1-e^{ia_n}|^2
\ . \ \ 
\label{f300}
\end{eqnarray}
One readily verifies that 
$|1-e^{ia}|=2|\sin(a/2)|\leq |a|$
for arbitrary $a\in \RR$,
yielding
\begin{eqnarray}
\langle\chi|\chi\rangle
\leq
\sum |c_n|^2 \,|a_n|^2\leq 
\max_n |a_n|^2
\ .
\label{f310}
\end{eqnarray}
By introducing (\ref{f310}) into (\ref{f270}) 
we can conclude
\begin{eqnarray}
|\delta_\psi| \leq  \da \max_n |a_n|
\ .
\label{f320}
\end{eqnarray}
Observing that this bound is independent 
of $|\psi\rangle$,
and taking into account
Eqs.~(\ref{f150}) and (\ref{f290}),
the announced final result 
(\ref{f140}) is recovered.


\begin{thebibliography}{99}
\bibitem{f5}
The variance $\sigma^2_v$ then refers to 
the band center.
\bibitem{f3}
While $t$ in (\ref{140}) is strictly speaking
considered as arbitrary but fixed,
one can show analogously as in 
Refs.~\cite{rei15} that a similar bound
also applies simultaneously for the vast
majority of all $t$'s 
within any preset time interval.
\bibitem{f4}
In principle, for certain combinations of $\rho(0)$ and
$A$ the condition (\ref{270}) may not be sufficient
to guarantee thermalization within the framework of
Refs.~\cite{deu91,rei15b}.
In practice, such cases have never been 
encountered and hence we tacitly ignore them.
\end{thebibliography}
\end{document}